\documentclass[prd, reprint, amsmath, amssymb, superscriptaddress, nofootinbib, aps, floatfix]{revtex4-2}
\usepackage{lmodern}

\usepackage[justification=justified,font=small,labelfont=bf]{caption}

\newcommand{\macro}[1]{\texttt{\textbackslash#1}}
\newcommand{\m}[1]{\macro{#1}}

\usepackage{textcase}
\usepackage{graphicx}%
\usepackage{multirow}
\usepackage{epstopdf}
\usepackage[utf8]{inputenc}
\usepackage[english]{babel}
\usepackage{footnote}
\usepackage{xcolor}
\usepackage{textcomp}
\usepackage{color} % for the command \textcolor
\usepackage{orcidlink}
\usepackage{soul}
\usepackage{siunitx}
\usepackage{comment}
\usepackage{subcaption}
\usepackage{threeparttablex} 
\usepackage[capitalise,noabbrev]{cleveref}
\usepackage{float}
\usepackage{threeparttable}
\usepackage[skip=6pt plus1pt, indent=12pt]{parskip}
\usepackage{silence}
\begin{document}

\title{Measurements of the neutron background for the Colorado Underground Research Institute (CURIE)}%

\author{Dakota K. Keblbeck \orcidlink{0000-0002-4087-9326}}\thanks{Corresponding author}
\affiliation{Department of Physics, Colorado School of Mines, Golden, Colorado 80401, USA}

\author{Uwe Greife \orcidlink{0000-0003-2906-8466}}
\affiliation{Department of Physics, Colorado School of Mines, Golden, Colorado 80401, USA}

\author{Eric Mayotte \orcidlink{0000-0003-2618-9166}}
\affiliation{Department of Physics, Colorado School of Mines, Golden, Colorado 80401, USA}

\author{Wouter Van De Pontseele \orcidlink{0000-0003-3422-3133}}
\affiliation{Department of Physics, Colorado School of Mines, Golden, Colorado 80401, USA}

\author{Kyle G. Leach \orcidlink{0000-0002-4751-1698}}
\affiliation{McDonald Institute for Astroparticle Physics, Queen's University, Kingston, Ontario, Canada}

%\author{Caitlyn Stone-Whitehead \orcidlink{0009-0000-5396-7328}}
%\affiliation{Department of Physics, Colorado School of Mines, Golden, Colorado 80401, USA}

%\author{Luke Wanner \orcidlink{0009-0006-2221-1740}}
%\affiliation{Department of Physics, Colorado School of Mines, Golden, Colorado 80401, USA}

%\author{Grace Wagner \orcidlink{0000-0003-0295-875X}}
%\affiliation{Department of Physics, Colorado School of Mines, Golden, Colorado 80401, USA}

\date{\today}%

\begin{abstract}
We present the first direct measurements of the neutron background at the Colorado Underground Research Institute (CURIE), a shallow-underground facility with an effective overburden of approximately $400$~m.w.e.\ in the Edgar Experimental Mine in Idaho Springs, Colorado. The radiogenic and muon-induced components were characterized in the Subatomic Particle Hideout using techniques matched to their respective energy scales. The radiogenic component was measured with a custom Bonner Cylinder Spectrometer comprising a $^{3}$He proportional counter and six cylindrical high-density polyethylene moderators. Its response functions were simulated in \textsc{geant4}, empirically calibrated using a $^{239}$Pu-$^{9}$Be source and the shadow-cone method, and then corrected to account for the approximately isotropic underground field using measurements at three detector orientations. Unfolding the measured count rates gives a total local scalar neutron flux, dominated by radiogenic production, for $E_n\lesssim\SI{10}{\MeV}$ of
$(2.51\pm0.58_{\rm stat}\pm0.39_{\rm sys})\times10^{-5}$~
\si{\per\square\centi\meter\per\second}, with the epithermal component accounting for approximately \SI{57}{\percent} of the integrated flux. The muon-induced component was measured with a BC-501A organic liquid scintillator operated for \SI{1431}{\hour}. Pulse shape discrimination identified 55~$\pm$~7 neutron candidates. Using the simulated detector response, these counts yield an integral muon-induced neutron flux at the rock--cavern boundary above \SI{10.5}{\MeV} of
$(2.09\pm0.29_{\rm stat}\,{}^{+0.79}_{-0.53}{}_{\rm sys})\times10^{-7}$~
\si{\per\square\centi\meter\per\second}. This flux uses the first-entry, boundary-area normalization of the previous CURIE simulation study. Its central value is approximately \SI{12}{\percent} higher than the site-specific prediction but is consistent within the reported uncertainties. Together, these measurements characterize the radiogenic and high-energy cosmogenic neutron environments at CURIE and provide quantitative input for planning future low-background experiments at the facility.
\end{abstract}

\maketitle

\section{Introduction}

Cosmic-ray interactions in the upper atmosphere generate cascades of secondary
particles, including pions and kaons~\cite{Kampert2012EASReview,globus2025ultrahighenergycosmic}. Muons produced primarily
through charged-pion and kaon decay constitute the most penetrating and abundant charged component of cosmic rays at
sea level~\cite{ParticleDataGroup:2026}. Their flux of approximately
\SI{170}{\per\square\meter\per\second} at sea level and mean energy of approximately
\SI{4}{\GeV}~\cite{mu_flux,ParticleDataGroup:2026} produce important backgrounds for experiments
searching for rare processes. Rock overburden provides the most effective passive suppression of
this component. As muons propagate from the surface to laboratory depth, they lose energy through
ionization and radiative processes, and their surviving flux decreases strongly with
depth~\cite{ParticleDataGroup:2026,Guo_2021}. Underground shielding is conventionally reported in
meters water equivalent (m.w.e.), permitting comparisons among facilities with different rock
compositions and overburden geometries. The deepest laboratories host flagship rare-event programs,
including direct dark-matter searches~\cite{lux-zeplin,delight-dm} and searches for neutrinoless
double-beta decay~\cite{nexo}, while shallower facilities support detector development, material
screening and radioassay~\cite{majorana_radio_assay,pnnl}, copper
electroforming~\cite{copper_electroforming}, and cryogenic and quantum-sensor
characterization~\cite{PhysRevB.110.024519}.

Reducing the cosmogenic muon flux does not eliminate it as a background source. Surviving muons
can often be identified through their characteristic ionization signals and rejected using
coincidence vetoes~\cite{Freund2016GerdaMuonVeto,THOMAS201347,formaggio_backgrounds}, but they also
produce secondary radiation in the material surrounding an underground laboratory. Neutrons are
particularly problematic because they can penetrate substantial shielding and interact differently
from the $\gamma$ rays for which high-$Z$ shielding is designed, requiring dedicated suppression
with hydrogen-rich moderators or neutron-absorbing materials. Their signatures can enter directly
into experimental regions of interest, e.g., $\gamma$ rays following inelastic scattering, $(n,n'\gamma)$, can deposit energy near a $0\nu\beta\beta$ endpoint~\cite{neutron_inelastic},
neutron capture can activate detector materials and target isotopes~\cite{Albert2018,Rogers_2020},
and elastic scattering can produce nuclear recoils that mimic dark-matter
interactions~\cite{Aprile_2013,mei_hime}. The underground neutron field has two physically distinct
origins with different energy scales and dependencies on site conditions, requiring both
components to be characterized.

The first origin is cosmogenic. Residual muons produce neutrons directly, through spallation
on nuclei and $\mu^{-}$ capture, and indirectly, through photons and hadrons generated in the
electromagnetic and hadronic cascades they induce along their
track~\cite{ParticleDataGroup:2026}. The resulting spectrum is hard, extending to
\SI{}{\GeV} scale~\cite{KUDRYAVTSEV2003688}, and these neutrons can traverse substantial shielding before reaching a detector. They are also among the least
constrained of all underground backgrounds. Predictions rely on muon photonuclear cross
sections carrying large uncertainties~\cite{KOKOULIN1999475, proposal, proposal2}, and
general-purpose transport codes are known both to disagree with one another and to fall below
measured yields, in some cases by factors of two or more~\cite{ARAUJO2005398, REICHHART201367,
sutanto_mu_n}. Direct measurements are scarce at any depth and scarcer still at shallow sites,
so each additional measurement serves the dual purpose of quantifying the background at a
given facility and constraining the codes used to predict it elsewhere.

The second origin is radiogenic. The $^{238}$U and $^{232}$Th decay chains in the surrounding
rock emit $\alpha$ particles of \SIrange{4}{9}{\MeV}, which produce neutrons through
($\alpha, n$) reactions on light nuclei; the spontaneous fission of $^{238}$U contributes a
smaller additional yield~\cite{alpha_n, spontaneous-fission, watt-spectrum}. This spectrum is
soft, and after moderation in the host rock it is confined to energies below
$\approx$~\SI{10}{\MeV}, where it substantially exceeds the muon-induced contribution at depths beyond approximately $100$~m.w.e.~\cite{alpha-n-depth, mei-alpha-n}. It is this
low-energy component that sets passive shielding and moderator requirements. Its
magnitude is governed by the local uranium and thorium content and by the abundance of
light-$Z$ target nuclei, and measured fluxes vary by an order of
magnitude between facilities of comparable depth~\cite{YOON2021102533, kamioka_3he,
lngs-alpha-n, canfranc-alpha-n}. Therefore, the radiogenic neutron flux cannot be reliably predicted by depth-based parameterizations, and it must
instead be measured in situ at each site.

The Colorado Underground Research Institute (CURIE) is a shallow-underground laboratory operated by the Colorado School of Mines within the Edgar Experimental Mine in Idaho Springs, Colorado~\cite{curie}. Excavated in stable gneiss and reached by a horizontal, rail-driven tunnel, the facility sits beneath an effective overburden of $\approx$~\SI{400}{}~m.w.e., attenuating the surface muon flux by a factor of 600--700~\cite{dakota}. Its primary research spaces are the Subatomic Particle Hideout (SPH) and Cryolab~I. The cosmogenic muon flux at CURIE has been measured and the underlying simulation framework validated in Ref.~\cite{dakota}, and the muon-induced secondary backgrounds have been predicted through coupled \textsc{mute}~\cite{mute, mute2} and \textsc{geant4}~\cite{g4} simulations in Ref.~\cite{dakota2}; the reader is referred to those works for a detailed description of the facility and its cosmogenic environment.

In this work, we present the first direct measurements of the neutron background at CURIE,
performed in the SPH. The radiogenic component below approximately \SI{10}{\MeV} was measured with
a custom Bonner Cylinder Spectrometer comprising a $^{3}$He proportional counter and cylindrical
high-density polyethylene moderators, and was unfolded into an energy spectrum and integrated flux.
The high-energy muon-induced component was measured with a BC-501A organic liquid scintillator,
which detects fast neutrons through nuclear recoils and separates them from the underground
$\gamma$-ray background using pulse shape discrimination. Together, these measurements characterize
the radiogenic and high-energy cosmogenic regimes and provide an experimental benchmark for the
simulation framework of Ref.~\cite{dakota2}.

\section{Radiogenic Neutrons}\label{sec:radio-neutrons}

The radiogenic neutron flux at CURIE was measured in the Subatomic Particle Hideout using a
custom Bonner Cylinder Spectrometer (BCS), consisting of a $^{3}$He proportional counter and
six cylindrical high-density polyethylene (HDPE) moderators. The detector response was simulated in \textsc{geant4} and
empirically calibrated against a $^{239}$Pu-$^{9}$Be source using the shadow-cone method, described in Section~\ref{sec:shadow-cone}. Because
the calibration measurements used parallel-beam geometries whereas the underground field is
approximately isotropic, an angular correction was derived before the measured count rates were
unfolded into the neutron spectrum and integrated flux.

\subsection{Bonner Cylinder Spectrometer}\label{sec:bcs}

The BCS exploits the \SI{5330}{\barn} thermal-neutron capture
cross section of $^{3}$He~\cite{Knoll2010}. The cylindrical
proportional counter used here~\cite{MaximusEnergy} was filled
with $^{3}$He at a pressure of \SI{2}{atm} and detects neutrons
through the reaction
\begin{equation}
    ^{3}\mathrm{He}(n,p)^{3}\mathrm{H},
\end{equation}
which releases \SI{764}{\keV} of kinetic energy, shared between
a \SI{573}{\keV} proton and a \SI{191}{\keV} triton. The resulting
pulse-height signature provides efficient neutron identification
with low sensitivity to the ambient $\gamma$-ray background.

Sensitivity above thermal energies is conventionally obtained by enclosing the counter in
spherical HDPE moderators of increasing diameter~\cite{bonner_sphere}. In the low neutron flux
at CURIE, however, the counter size permitted by a spherical moderator would require
prohibitively long acquisition times. We therefore use cylindrical moderators, which accommodate
a larger active volume at the cost of an angularly dependent response. The correction for this
dependence is described in Sec.~\ref{sec:isotropic-corrections}.

\begin{figure}[!t]
    \centering
    \includegraphics[width=\linewidth]{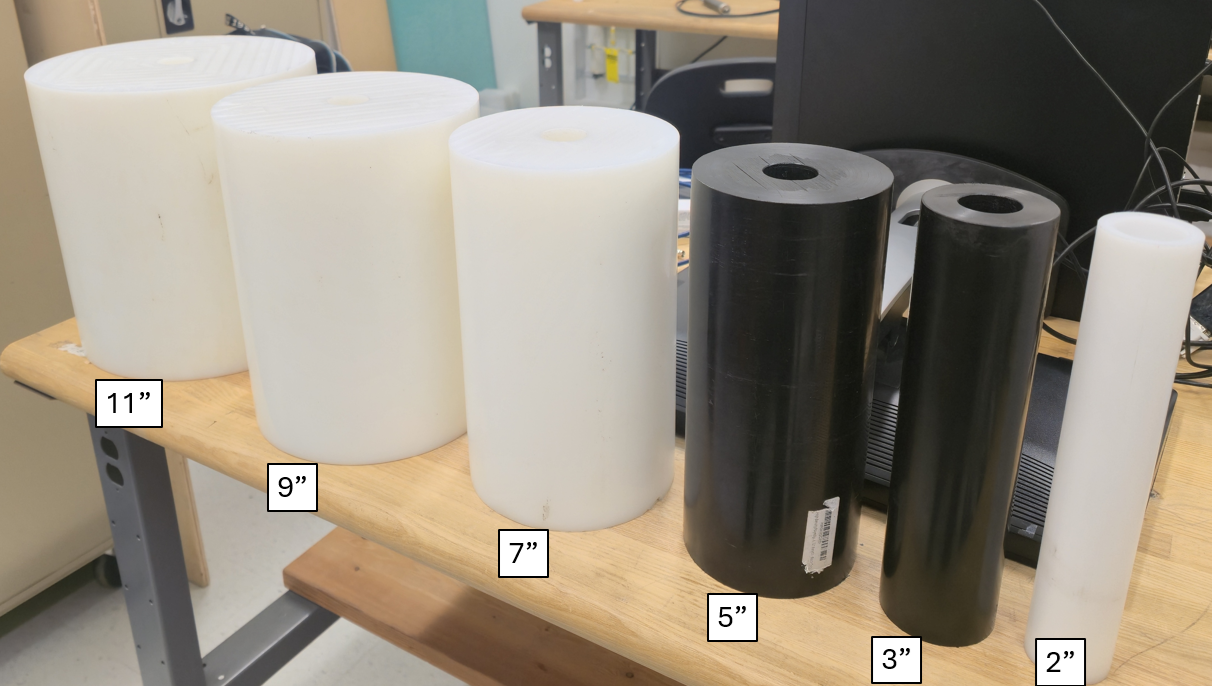}
    \caption{Complete set of the six HDPE cylindrical moderators.}
    \label{fig:spectrometer_set}
\end{figure}

Six HDPE cylinders of common height \SI{12}{in} and diameters from \SI{2}{} to
\SI{11}{in} were selected using \textsc{geant4} response simulations. Together with the bare
counter, they provide seven configurations spanning thermal energies to approximately
\SI{40}{\MeV}, as shown in \cref{fig:spectrometer_set} and summarized in
\cref{tab:spectrometer-configs}. Each cylinder was capped with an HDPE plug, which fully covered the exposed $^{3}$He proportional counter.

\begin{table}[!b]
\caption{Detector configurations for the Bonner Cylinder Spectrometer, with the moderator dimensions given as diameter $\times$ height $\times$ wall thickness, and the peak sensitivity of the simulated neutron energies.}
\label{tab:spectrometer-configs}
\renewcommand{\arraystretch}{1.35}
\bgroup % Keeps the change local to this group
\setlength{\tabcolsep}{8pt} % Increase padding (Default is 6pt)
%\begin{ruledtabular}
\begin{tabular}{l c c}
Configuration & $D \times H \times T$ (in) & $E_{\rm peak}$ (MeV) \\
\colrule
Bare & $1.3 \times 8.9$ $\times$ -- & Thermal \\
Cylinder 1 (C1) & $2.0 \times 12.0 \times 0.35$ & $10^{-7}$ \\
Cylinder 2 (C2) & $3.0 \times 12.0 \times 0.85$ & $10^{-6}$ \\
Cylinder 3 (C3) & $5.0 \times 12.0 \times 1.85$ & 1 \\
Cylinder 4 (C4) & $7.0 \times 12.0 \times 2.85$ & 3 \\
Cylinder 5 (C5) & $9.0 \times 12.0 \times 3.85$ & 5 \\
Cylinder 6 (C6) & $11.0 \times 12.0 \times 4.85$ & 10 \\
\end{tabular}
\egroup
%\end{ruledtabular}
\end{table}

\subsection{Response Functions}\label{sec:response-functions}

The response function, $R(E)$, relates an incident neutron field to the resulting detector signal,
incorporating the detector geometry, material composition, interaction cross sections, and
intrinsic detection efficiency. For a monoenergetic neutron fluence, it is defined as
\begin{equation}
    R(E) = \frac{N_d}{\Phi(E)}\,,
    \label{eq:response-function}
\end{equation}
where $N_d$ is the number of detected $^{3}\mathrm{He}(n,p)^{3}\mathrm{H}$ events and
$\Phi(E)$ is the incident fluence. For isotropic incidence, this is the local scalar fluence, integrated over all arrival directions. The response therefore has units of area and can be
interpreted as an effective response area that combines the geometric acceptance and detection
probability at neutron energy $E$. Increasing the moderator thickness shifts the detector
sensitivity toward higher neutron energies.

All response functions were simulated in \textsc{geant4}~\cite{g4} version~11.2.1
using the \textsc{shielding\_hp} physics list~\cite{geant4_Shielding}
and G4NDL neutron data. Thermal scattering for hydrogen bound in
polyethylene was enabled. The $^{3}$He counter was modeled according
to the manufacturer specifications~\cite{MaximusEnergy}, and the
moderators were modeled as HDPE with a density of
\SI{0.96}{\gram\per\cubic\centi\meter} and mass fractions of
\SI{14.37}{\percent} hydrogen and \SI{85.63}{\percent} carbon.
Each response function was evaluated at 140 logarithmically spaced
neutron energies from \SI{1e-9}{} to \SI{40}{\MeV}, with
\num{1e8} neutrons simulated at each energy.

For a parallel beam, the incident fluence is defined by the number of generated neutrons,
$N_{\rm gen}$, and the area of the generation plane, $A_{\rm gen}$. Equation~\ref{eq:response-function}
then becomes
\begin{equation}
    R(E) = \frac{N_d}{N_{\rm gen}/A_{\rm gen}}
         = \frac{N_d A_{\rm gen}}{N_{\rm gen}}\,.
    \label{eq:parallel-rf}
\end{equation}
For incidence parallel to the cylinder axis, the generation plane is
circular, whereas for side-on incidence it is rectangular, with
$A_{\rm gen}=DH$, where $D$ and $H$ are the cylinder diameter and
height. At an intermediate incidence angle $\theta_{\rm inc}$,
measured relative to the cylinder axis, the projected area is
\begin{equation}
    A_{\rm proj,c}
    = \pi\left(\frac{D}{2}\right)^2\cos\theta_{\rm inc}
    + DH\sin\theta_{\rm inc}\,.
\end{equation}
This expression recovers the face-on and side-on projected areas at
$\theta_{\rm inc}=\SI{0}{\degree}$ and
$\theta_{\rm inc}=\SI{90}{\degree}$, respectively.

Previous cylindrical neutron spectrometers have generally calculated their response functions
using parallel-beam simulations~\cite{rf_scaling1,bonner-cylinder1,bonner-cylinder2}. This
treatment is appropriate for the source characterization measurements described in
Sec.~\ref{sec:characterization}, but not for the approximately isotropic neutron field encountered
underground. Applying a response calculated for a single incident direction to an isotropic field
can therefore introduce a systematic bias~\cite{rf_scaling1}.

For an isotropic field, no unique generation plane exists. The appropriate geometric quantity is
instead the mean projected area presented by the detector over all incident directions, which we
denote the effective acceptance area, $A_{\rm eff}$. For a convex body, Cauchy's
mean-projected-area theorem gives
\begin{equation}
    A_{\rm eff} = \frac{S}{4}\,,
\end{equation}
where $S$ is the total surface area~\cite{cauchy}.

The effective acceptance areas were also evaluated directly in \textsc{geant4}. Each detector
configuration was placed at the center of a spherical source surface, with neutron starting
positions sampled uniformly over the surface and incident directions sampled from a Lambertian
distribution. This produces a spatially uniform, angularly isotropic fluence within the sphere.
If $R_s$ is the radius of the source sphere and $N_{\rm hit}$ is the number of generated
neutrons whose trajectories intersect the detector assembly, then
\begin{equation}
    A_{\rm eff}
    = \pi R_s^2\frac{N_{\rm hit}}{N_{\rm gen}}\,.
    \label{eq:effective-area-isotropic}
\end{equation}
The isotropic response function is consequently
\begin{equation}
    R(E)
    = \frac{N_d A_{\rm eff}}{N_{\rm hit}}\,.
    \label{eq:iso-rf}
\end{equation}
The simulated effective areas, listed in \cref{tab:eff-areas}, agree with the corresponding
analytic values of $S/4$, confirming the normalization of the isotropic source geometry.

\begin{table}[ht]
\caption{Effective acceptance areas for an isotropic neutron field, calculated for each detector
configuration.}
\label{tab:eff-areas}
\renewcommand{\arraystretch}{1.35}
\begin{tabular}{lc}
Name & $A_{\rm eff}$ (\si{\square\centi\meter}) \\
\colrule
Bare & 65.94 \\
C1 & 131.74 \\
C2 & 205.22 \\
C3 & 367.36 \\
C4 & 549.78 \\
C5 & 752.46 \\
C6 & 975.41 \\
\end{tabular}
\end{table}

\begin{figure}[ht]
    \centering
    \includegraphics[width=\linewidth]{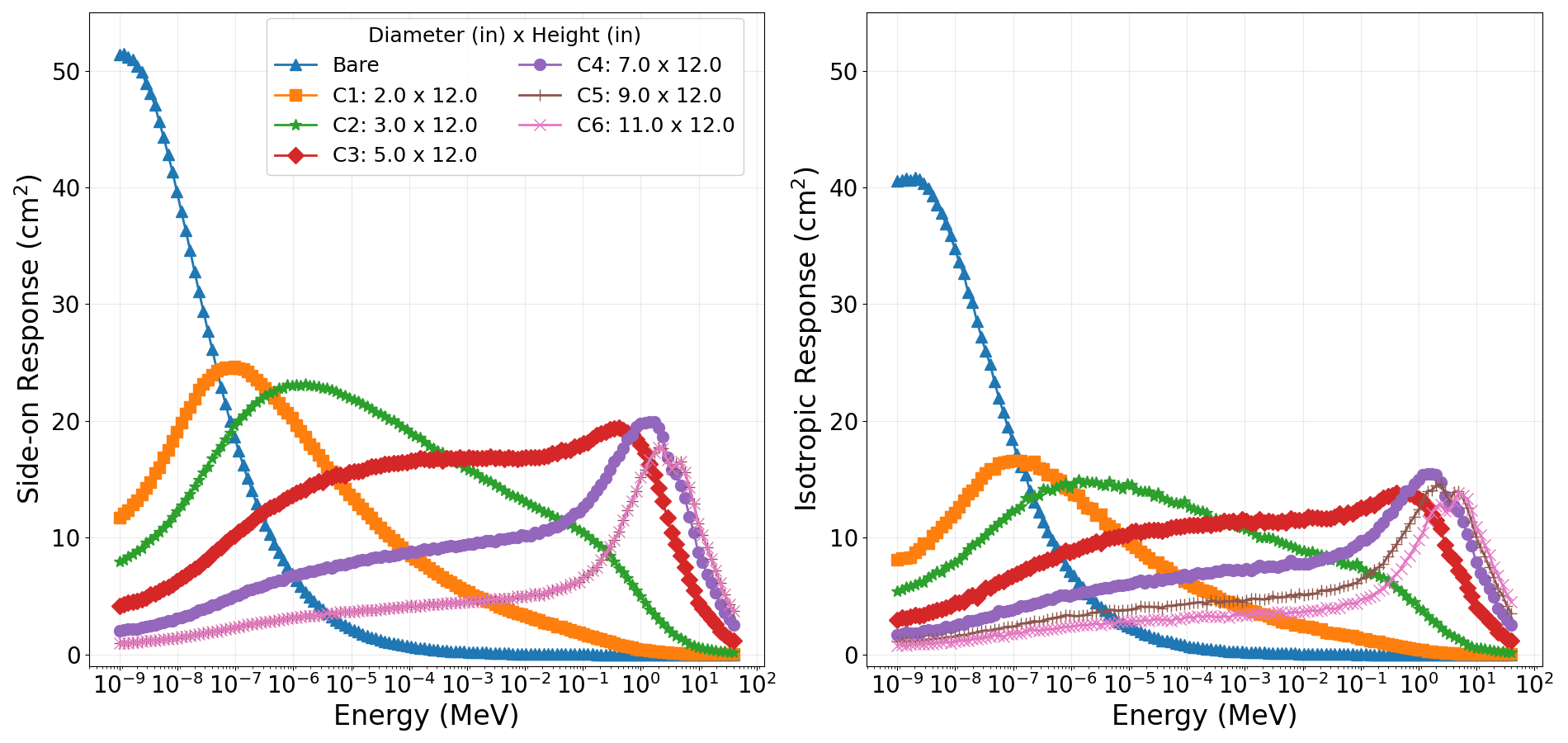}
    \caption{Simulated response functions for the seven BCS configurations for side-on
    parallel-beam incidence (left) and isotropic incidence (right).}
    \label{fig:side-v-iso-rf}
\end{figure}

The isotropic response functions retain the general energy dependence of their side-on
counterparts but are lower in magnitude, as shown in \cref{fig:side-v-iso-rf}. This difference
reflects the angular dependence of the cylindrical geometry and demonstrates the importance of
calculating the response for the incident field being measured.

\subsection{Detector Characterization}\label{sec:characterization}

\begin{figure}[ht]
    \centering
    \includegraphics[width=\linewidth]{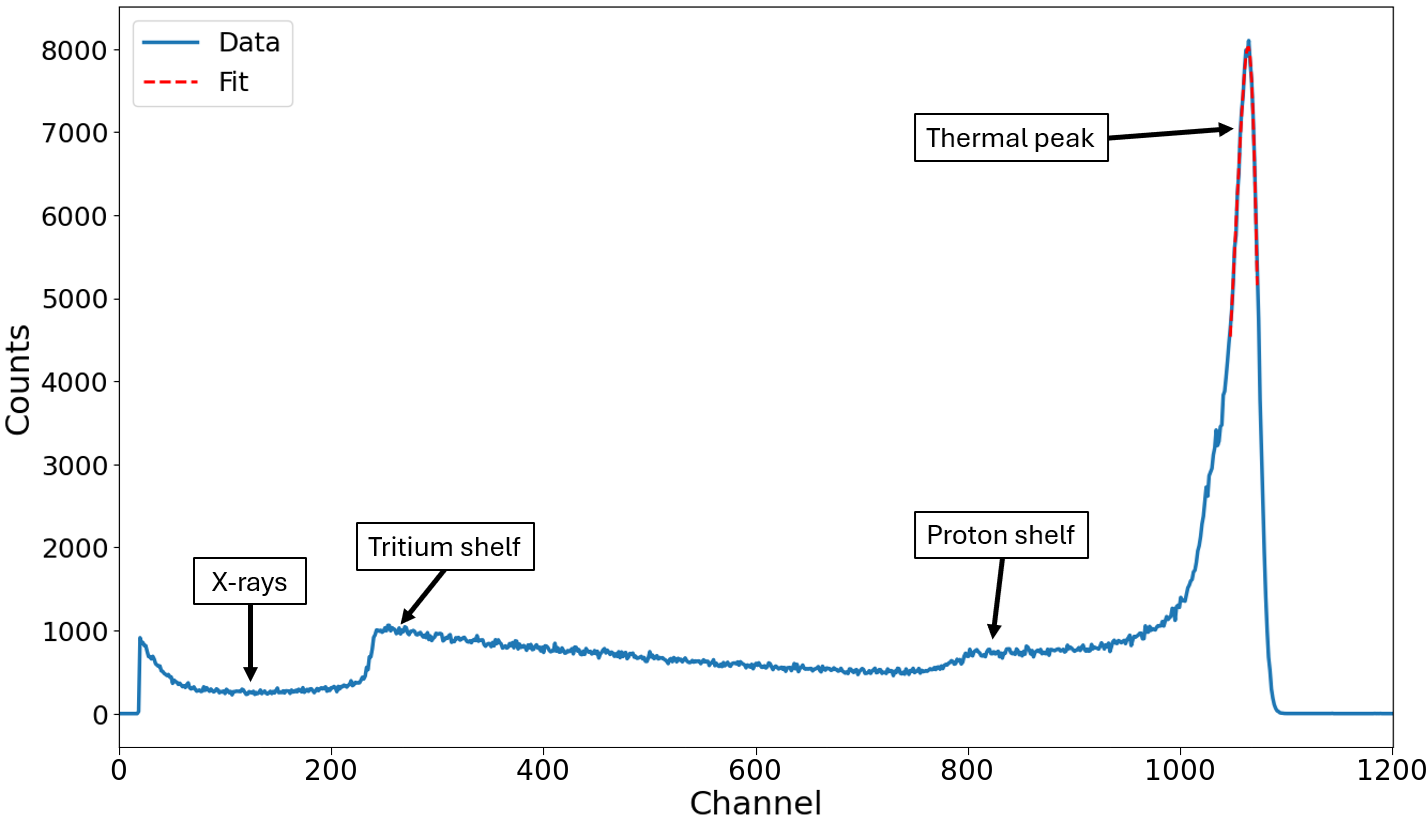}
    \caption{Pulse-height spectrum acquired with the bare $^{3}$He counter, showing the
    full-energy thermal-neutron capture peak and the proton and triton shelves. The fit to the
    capture peak determines the channel-to-energy calibration used to define the neutron ROI.}
    \label{3he_spectrum}
\end{figure}

As mentioned earlier, the $^{3}\mathrm{He}(n,p)^{3}\mathrm{H}$ reaction releases kinetic energy shared between a \SI{573}{\keV} proton and a \SI{191}{\keV} triton. When both reaction products stop within the
active gas, their full combined energy is recorded. Events in which one product reaches the
counter wall before depositing its full energy form the proton and triton shelves below the
full-energy peak~\cite{Knoll2010}. These characteristic features are visible in the bare-counter
pulse-height spectrum shown in \cref{3he_spectrum}.

The full-energy peak was used to determine the channel-to-energy calibration. Neutron events were
then selected using the region of interest (ROI) from channel 245 to channel 1085. The lower
boundary corresponds to the \SI{191}{\keV} triton edge, while the upper boundary lies above the
\SI{764}{\keV} full-energy peak to ensure its complete inclusion. This ROI retains the
$^{3}\mathrm{He}(n,p)^{3}\mathrm{H}$ events while rejecting the lower-energy region dominated
by ambient $\gamma$ rays and electronic noise.

Potential background contributions from intrinsic alpha
activity in the $^{3}$He proportional counter were also
investigated. A six-hour background measurement was
performed in the CoorsTek basement with the bare
counter surrounded by \SI{6}{in} of HDPE to suppress
the ambient neutron background. No events were observed
within the neutron ROI, indicating no measurable
intrinsic alpha contribution under these conditions.

The simulated response functions were characterized experimentally using the shadow-cone
method~\cite{n_spectrometer_cal1,n_spectrometer_cal2}. This method separates neutrons arriving
directly from a calibration source from those scattered by the air, floor, walls, and surrounding
materials.

\subsubsection{Shadow Cone Method}\label{sec:shadow-cone}

The shadow cone consists of eight \SI{1}{in} thick steel slabs,
with an Fe content greater than \SI{95}{\percent}, followed by
twelve \SI{1}{inch} thick slabs of \SI{5}{\percent} borated
polyethylene. The assembly approximates a pyramidal
frustum with a total length of \SI{20}{in}, following International
Organization for Standardization guidelines for shadow-cone
measurements~\cite{n_spectrometer_cal2,iso-sc3}. The smaller,
source-facing end measures \SI{2.75}{in}$\times$\SI{1.38}{in},
and the larger, detector-facing end measures
\SI{6.23}{in}$\times$\SI{5.07}{in}. The steel section scatters
and degrades the energy of neutrons in the direct beam, while
the borated polyethylene moderates and absorbs the remaining
neutrons.

Two measurements are required for each detector configuration. A measurement without the cone
contains both direct and scattered neutrons together with the ambient background. With the cone, the direct beam is blocked, leaving the scattered
contribution and ambient background. The ambient background is
assumed to remain unchanged between the two measurements and
therefore cancels in their subtraction. The number of direct neutron events is therefore
\begin{equation}
    N_{\rm direct}
    = \left(N_{\rm total}-N_{\rm scattered}\right)F_{\rm A}(d)\,,
    \label{eq:direct-counts}
\end{equation}
where $F_{\rm A}(d)$ corrects for neutrons scattered out of the direct beam by the air over the
source-to-detector distance $d$. This correction is
\begin{equation}
    F_{\rm A}(d)=e^{\overline{\Sigma}d}\,,
    \label{eq:air-correction}
\end{equation}
where $\overline{\Sigma}$ is the spectrum-averaged macroscopic neutron cross section in air.
Subtracting the shadow-cone measurement and applying this correction isolates the direct-beam
response needed for comparison with the simulated response functions.

\subsubsection{PuBe Neutron Source}\label{sec:pube-source}

The characterization measurements used a cylindrical, doubly encapsulated
$^{239}$Pu-$^{9}$Be source (NUMEC-B\footnote{Nuclear Materials and Equipment Corporation}, \SI{32}{\gram} Pu, \SI{2}{Ci}) with a reported neutron
emission rate of approximately \SI{3.6e6}{}~n/s. The long $^{239}$Pu half-life of approximately
\SI{24000}{yr} makes the neutron emission rate effectively constant over the timescale of the
measurements. Neutrons are produced through
\begin{equation}
\begin{aligned}
{}^{239}\mathrm{Pu} &\rightarrow {}^{235}\mathrm{U} + \alpha \,, \\
\alpha + {}^{9}\mathrm{Be} &\rightarrow {}^{12}\mathrm{C} + n + \gamma.
\end{aligned}
\end{equation}

The source was modeled in \textsc{geant4} using its reported
plutonium activity and the double-encapsulation geometry of
Ref.~\cite{pube-capsule}. The simulated neutron emission rate
was \SI{2.1}{\percent} higher than the reported value.
The simulated mean neutron energies are
$(4.06^{+0.08}_{-0.04})$~\si{\MeV} over
\SIrange{0}{10.5}{\MeV} and
$(5.15^{+0.04}_{-0.01})$~\si{\MeV} over
\SIrange{1}{14.1}{\MeV}, where the uncertainties arise from
the energy binning. These mean energies agree with measurements
of comparable PuBe sources~\cite{leone_pube,SODERSTROM_pube}.

\subsubsection{Measurements}

\begin{figure}[ht]
    \centering
    \includegraphics[width=\linewidth]{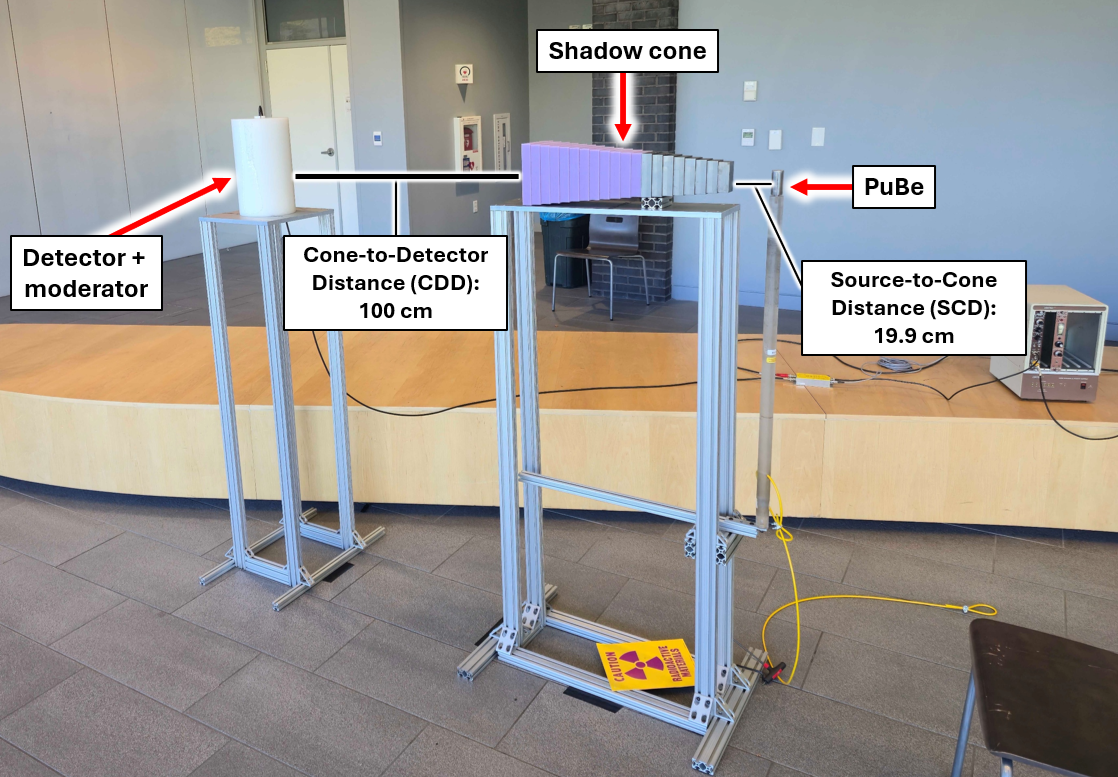}
    \caption{Shadow-cone configuration used for the side-on PuBe characterization measurements.}
    \label{fig:pube_setup}
\end{figure}

The characterization measurements were performed in the CoorsTek Atrium at the Colorado School
of Mines, a large open space selected to reduce the contribution from room-return neutrons. The
detector was initially oriented side-on to the source, as shown in \cref{fig:pube_setup}. The
cone-to-detector distance was fixed at \SI{100}{\centi\meter} following shadow-cone measurement
guidelines~\cite{iso-sc2,iso-sc3}.

The source-to-cone distance was varied from \SI{4.0}{} to
\SI{39.5}{\centi\meter} using the C4 configuration to identify
a plateau region in the measured cone-in count
rate~\cite{n_spectrometer_cal2,iso-sc2}. A source-to-cone distance of
\SI{19.9}{\centi\meter} was selected within this region to
fully shadow the detector while limiting suppression of
the scattered field. This distance was used for all seven
detector configurations.

Each detector configuration was measured for \SI{300}{\second} with and without the shadow cone.
The air-outscatter correction of Eq.~\ref{eq:air-correction} was evaluated over the PuBe neutron
spectrum, giving
$\overline{\Sigma}=\SI{8.35e-5}{\per\centi\meter}$. Corresponding \textsc{geant4}
simulations were performed using the PuBe spectrum and the same source-detector geometry, with
air and room-return contributions omitted, providing only the response to a direct neutron flux. Each simulation corresponded to an equivalent live
time of \SI{1200}{\second}.

For each configuration, the ratio of the measured to simulated direct count rates defines an
empirical scaling factor,
\begin{equation}
    S_i = \frac{R_{{\rm meas},i}}{R_{{\rm sim},i}}\,.
    \label{eq:side-scaling-factor}
\end{equation}
These factors empirically normalize the simulated response functions
to the PuBe measurements~\cite{rf_scaling1,rf_scaling2,rf_scaling3,
rf_scaling4,rf_scaling5}. Each factor is applied uniformly over
neutron energy, retaining the simulated energy dependence of the
response. This procedure assumes that the residual discrepancy
can be represented by an overall normalization correction.
Because the calibration measures the response integrated over
the PuBe spectrum, it does not independently constrain
energy-dependent discrepancies. The measured and simulated rates
and the resulting side-on scaling factors are listed in
\cref{tab:bcs-calibration-summary}.

\begin{table}[ht]
\caption{Measured and simulated PuBe count rates and the resulting side-on response-function scaling factors. The reported uncertainties are the quadrature sums of the statistical and systematic contributions, propagated separately for the upper and lower bounds.}
\label{tab:bcs-calibration-summary}
\renewcommand{\arraystretch}{1.35}
\bgroup
\setlength{\tabcolsep}{8pt}
\begin{tabular}{lccc}
Name & $R_{\rm meas}$ (cps) & $R_{\rm sim}$ (cps) & $S$ \\
\colrule
Bare & $0.06^{+0.02}_{-0.05}$ & $0.07^{+0.01}_{-0.02}$ & $0.83^{+0.36}_{-0.70}$ \\
C1 & $1.84^{+0.10}_{-0.12}$ & $3.66^{+0.16}_{-0.18}$ & $0.50^{+0.04}_{-0.04}$ \\
C2 & $31.73^{+1.08}_{-1.07}$ & $32.09^{+1.02}_{-0.61}$ & $0.99^{+0.04}_{-0.05}$ \\
C3 & $127.19^{+4.18}_{-4.04}$ & $113.11^{+2.86}_{-3.36}$ & $1.12^{+0.05}_{-0.04}$ \\
C4 & $155.03^{+5.03}_{-5.01}$ & $146.95^{+4.71}_{-4.77}$ & $1.05^{+0.05}_{-0.05}$ \\
C5 & $139.29^{+4.51}_{-4.58}$ & $140.13^{+4.99}_{-4.12}$ & $0.99^{+0.04}_{-0.05}$ \\
C6 & $112.89^{+3.67}_{-3.65}$ & $116.69^{+2.56}_{-5.38}$ & $0.97^{+0.05}_{-0.04}$ \\
\end{tabular}
\egroup
\end{table}

\subsubsection{Spectrum Unfolding}\label{sec:spectra-unfolding}

Each BCS configuration measures the incident neutron field through a broad, energy-dependent
response. Expressed in terms of count rate and differential neutron flux, the measured rate for
configuration $j$ is
\begin{equation}
    C_j
    = \int R_j(E)\frac{d\Phi(E)}{dE}\,dE\,,
    \label{eq:counts-integral}
\end{equation}
where $R_j(E)$ is the response function and $d\Phi/dE$ is the differential neutron flux. The
equivalent count--fluence formulation is obtained by multiplying the count rate and flux by the
measurement live time. Discretizing Eq.~\ref{eq:counts-integral} over the response-function energy
grid gives
\begin{equation}
    C_j
    = \sum_{i=1}^{n}
      R_{j,i}\,\phi_i\,\Delta E_i\,,
    \label{eq:counts-sum}
\end{equation}
where $\phi_i=(d\Phi/dE)_i$ and $\Delta E_i$ is the energy bin width.

The seven measured responses cannot uniquely determine the neutron flux in 140 energy bins.
Moreover, the response functions are broad and strongly overlapping, so additional moderator
configurations would provide only partially independent information. The measurement therefore
constrains broad thermal, epithermal, and fast components more reliably than fine bin-to-bin
structure. Spectrum unfolding regularizes this underdetermined inversion by imposing physical
constraints and incorporating an input spectrum that guides the solution where the data provide
limited information.

Three established unfolding algorithms are used in this work: \textsc{maxed}, \textsc{gravel}, and \textsc{mlem}. \textsc{maxed} maximizes the
entropy of the solution relative to a supplied default spectrum while constraining the agreement
between the predicted and measured count rates through a $\chi^2$ condition~\cite{MAXED1,MAXED2}.
The \textsc{gravel} and maximum-likelihood expectation-maximization (\textsc{mlem})
algorithms instead iteratively update an initial spectrum until its response-folded count rates
converge toward the measurements~\cite{gravel,mlem}.

The PuBe data provide an internal closure test of the calibrated response
matrix and unfolding procedure. The simulated PuBe spectrum described in
Sec.~\ref{sec:pube-source} was normalized and supplied as the initial
spectrum on the same 140-bin energy grid as the response functions.
Both \textsc{gravel} and \textsc{mlem} reproduce its principal features,
with \textsc{gravel} providing better agreement below approximately
\SI{0.1}{\MeV}. Because the measured rates also determine the response
scaling factors and the source spectrum supplies the initial shape,
this agreement establishes internal consistency but does not independently
validate the energy dependence of the response functions.
For the underground analysis, \textsc{maxed} is additionally included
to provide a reconstruction based on a methodologically distinct
regularization procedure.

\begin{figure}[!htb]
    \centering
    \includegraphics[width=\linewidth]{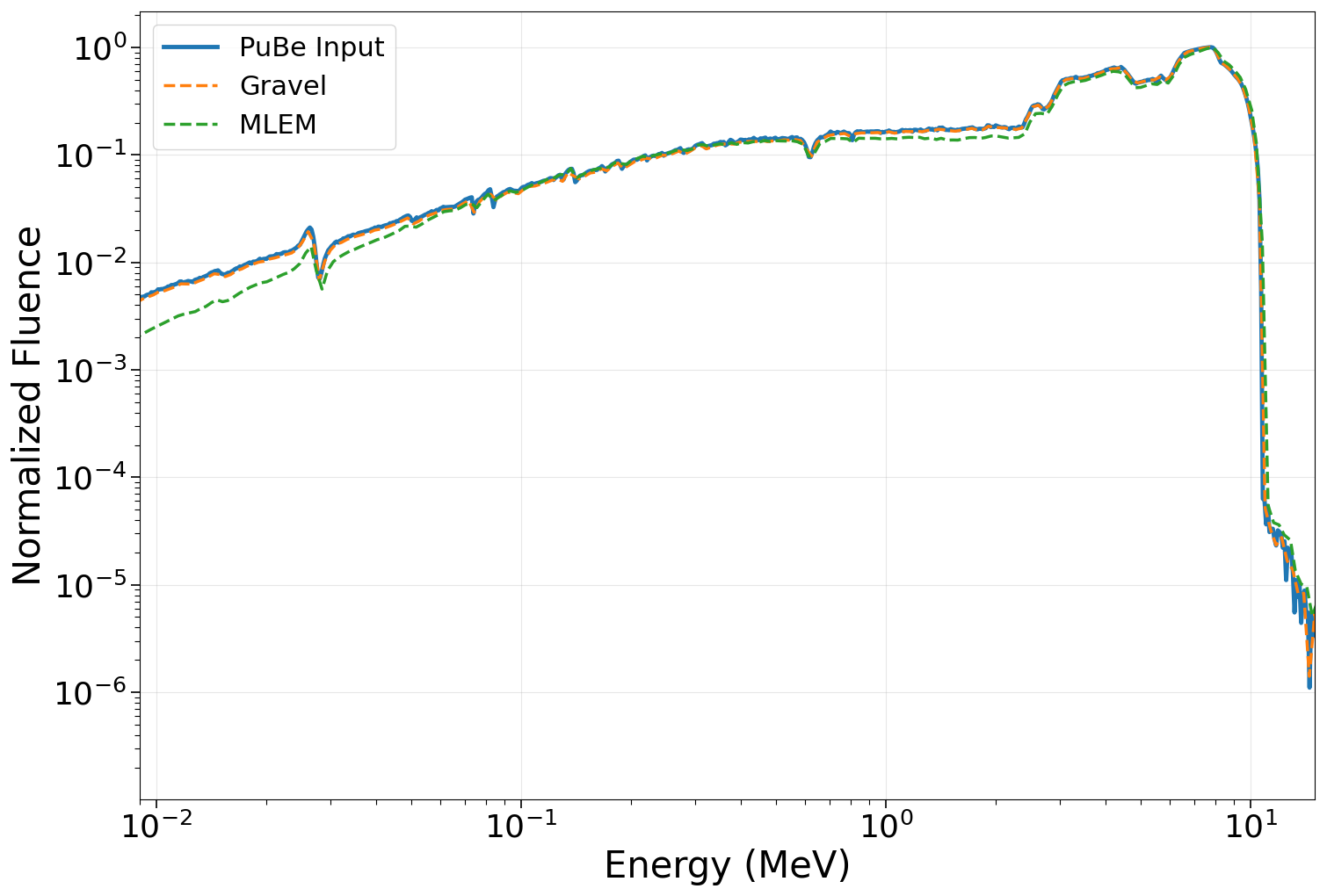}
    \caption{PuBe neutron spectra reconstructed using \textsc{gravel}
and \textsc{mlem}, compared with the simulated PuBe spectrum used
as the initial spectrum. The spectra are normalized to make
deviations in spectral shape more visible.}
    \label{fig:gravel-v-mlem}
\end{figure}

\begin{figure}[ht]
    \centering
    \includegraphics[width=\linewidth]{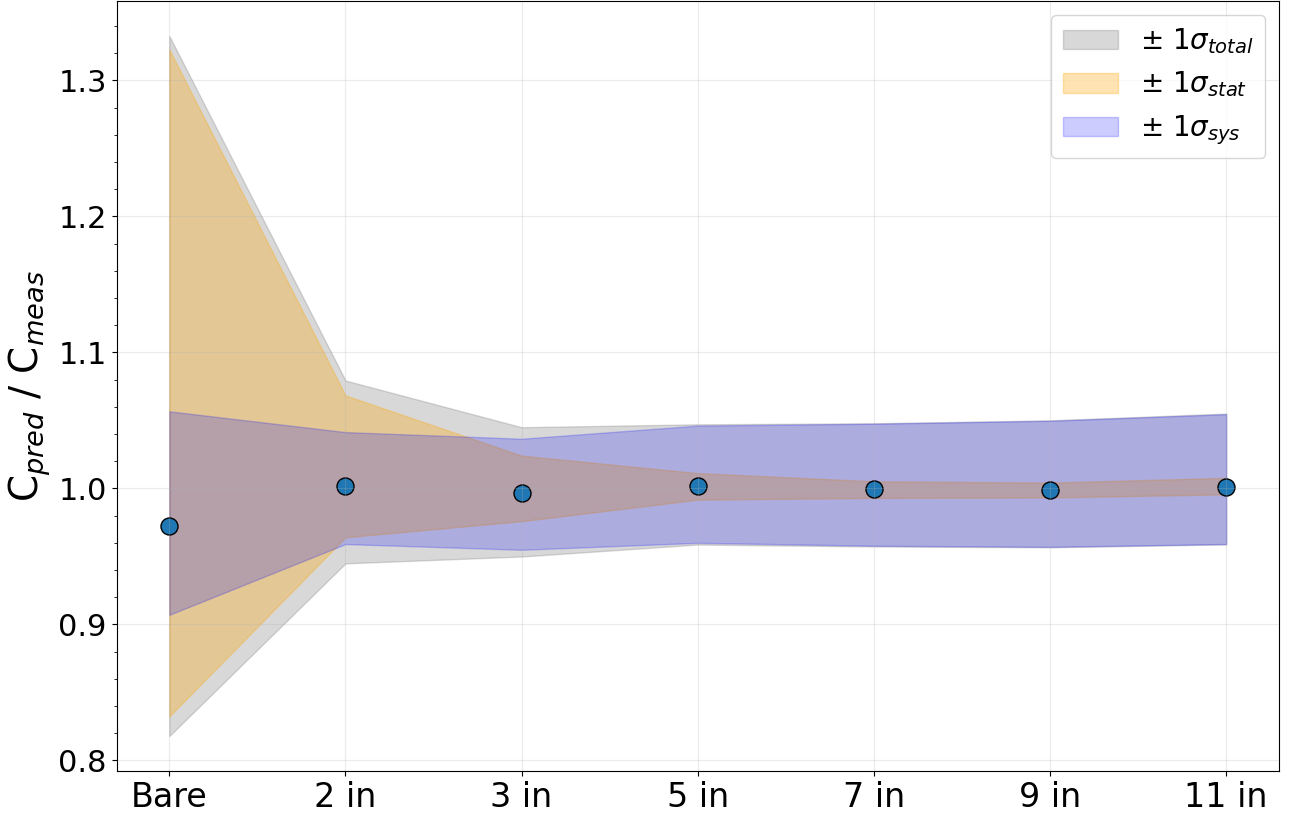}
    \caption{Ratio of the predicted to measured PuBe count rates for each detector
    configuration, obtained by refolding the \textsc{gravel} reconstruction through the
    calibrated response matrix. Statistical and systematic uncertainties are shown separately,
    and the total uncertainty is their quadrature sum.}
    \label{fig:unfolding-results}
\end{figure}

Each unfolded spectrum is evaluated by refolding it through the response matrix,
\begin{equation}
    C_{{\rm pred},j}
    = \sum_{i=1}^{n}
      R_{j,i}\,\phi_{{\rm pred},i}\,\Delta E_i\,,
    \label{eq:pred-counts}
\end{equation}
and comparing the predicted and measured count rates~\cite{n-spec-unfolding1,Stuenkel01071999}.
For the \textsc{gravel} reconstruction, the predicted rates agree with all seven measured rates
within their uncertainties, as shown in \cref{fig:unfolding-results}. Integrating the unfolded
spectrum gives a source emission rate of
\begin{equation}
    S_n =
    (3.66 \pm 0.04_{\rm stat} \pm 0.18_{\rm sys})
    \times 10^{6}~\mathrm{n/s}\,,
\end{equation}
consistent with the reported PuBe emission rate. The recovered spectral shape, source strength,
and response-folded count rates demonstrate closure of the calibrated response and unfolding
procedure before it is applied to the underground measurements.

\subsection{Isotropic Corrections}\label{sec:isotropic-corrections}

The scaling factors derived in Sec.~\ref{sec:characterization} were measured in a side-on
parallel-beam geometry, whereas the underground neutron field is assumed to be approximately
isotropic. Ideally, the corrections applied to the isotropic response functions would themselves
be determined using a known isotropic neutron field. No calibration source, however, produces a
well-characterized and truly isotropic field over the required energy range. This limitation has
generally led experiments either to use spherical moderators, whose response is intrinsically
near-isotropic, or to approximate the isotropic response using parallel-beam
measurements~\cite{rf_scaling1,bonner-cylinder1,bonner-cylinder2,neutron_shield,YOON2021102533}.

\begin{figure}[ht]
    \centering
    \includegraphics[width=\linewidth]{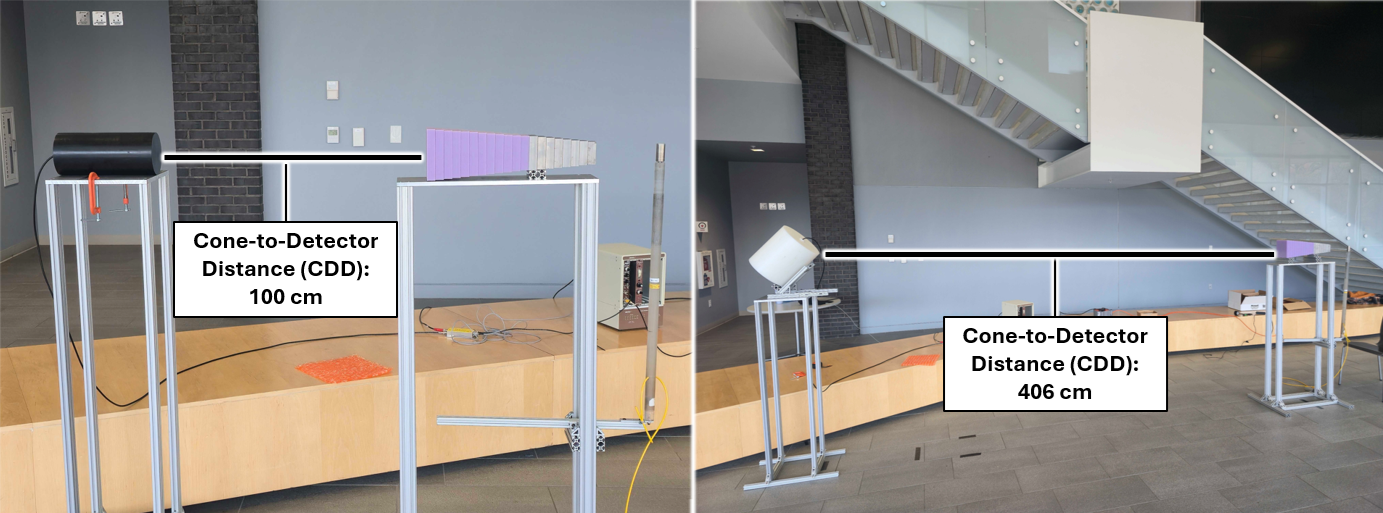}
    \caption{Experimental shadow-cone configurations used for the
    \SI{90}{\degree} (left) and \SI{45}{\degree} (right) detector orientations.}
    \label{fig:pube-rotations}
\end{figure}

To account for the angular response of the cylindrical moderators, additional shadow-cone
measurements were performed with the detector rotated by \SI{45}{\degree} and
\SI{90}{\degree} relative to the original side-on orientation
(\cref{fig:pube-rotations}). Together with the \SI{0}{\degree} measurements of
Sec.~\ref{sec:characterization}, these measurements provide scaling factors
$S(\SI{0}{\degree})$, $S(\SI{45}{\degree})$, and $S(\SI{90}{\degree})$ for each
configuration, spanning the side-on, intermediate, and face-on geometries.

The cone-to-detector distance was selected independently for each orientation so that the near
face of the cone fully shadowed the projected area of the largest moderator, C6. Distances of
\SI{406}{\centi\meter} and \SI{100}{\centi\meter} were used for the
\SI{45}{\degree} and \SI{90}{\degree} measurements, respectively. The larger distance at
\SI{45}{\degree} was required by the greater projected area in that orientation. These
differences do not bias the extracted scaling factors because each measurement was compared with
a \textsc{geant4} simulation performed using the same source-detector geometry. Consequently,
each ratio
\begin{equation}
    S(\theta)
    = \frac{R_{\rm meas}(\theta)}{R_{\rm sim}(\theta)}
\end{equation}
is determined at a matched source-detector distance and can be combined with those from the
other orientations.

Each measured scaling factor describes the detector response for a single incident orientation,
whereas an isotropic field illuminates the detector from all directions. The correction required
for the isotropic response functions is therefore the response-weighted average
\begin{equation}
    S_{\rm iso}
    =
    \frac{
        \displaystyle
        \int_{0}^{\pi/2} S(\theta)w(\theta)\,d\theta
    }{
        \displaystyle
        \int_{0}^{\pi/2} w(\theta)\,d\theta
    }\,,
    \label{eq:siso-continuous}
\end{equation}
where $w(\theta)$ accounts for both the angular phase space of an isotropic field and the
detector response at each orientation.

For an isotropic neutron field, the angular weighting includes
the solid-angle element $\sin\theta_{\rm inc}\,d\theta_{\rm inc}$,
where $\theta_{\rm inc}$ is measured relative to the cylinder
axis. The detector's angular sensitivity is represented by
$R_{\rm sim}(\theta)$, the simulated count rate integrated over
the PuBe neutron spectrum. These rates were calculated using
a common source-detector geometry and source normalization,
so their angular variation incorporates the projected area
and intrinsic detection efficiency averaged over the PuBe
spectrum. The common normalization cancels in
Eq.~\ref{eq:siso-continuous}. The angular weight is therefore
\begin{equation}
    w(\theta)
    = R_{\rm sim}(\theta)\sin\theta_{\rm inc}\,.
    \label{eq:iso-weight}
\end{equation}

\begin{figure}[ht]
    \centering
    \includegraphics[width=0.8\linewidth]{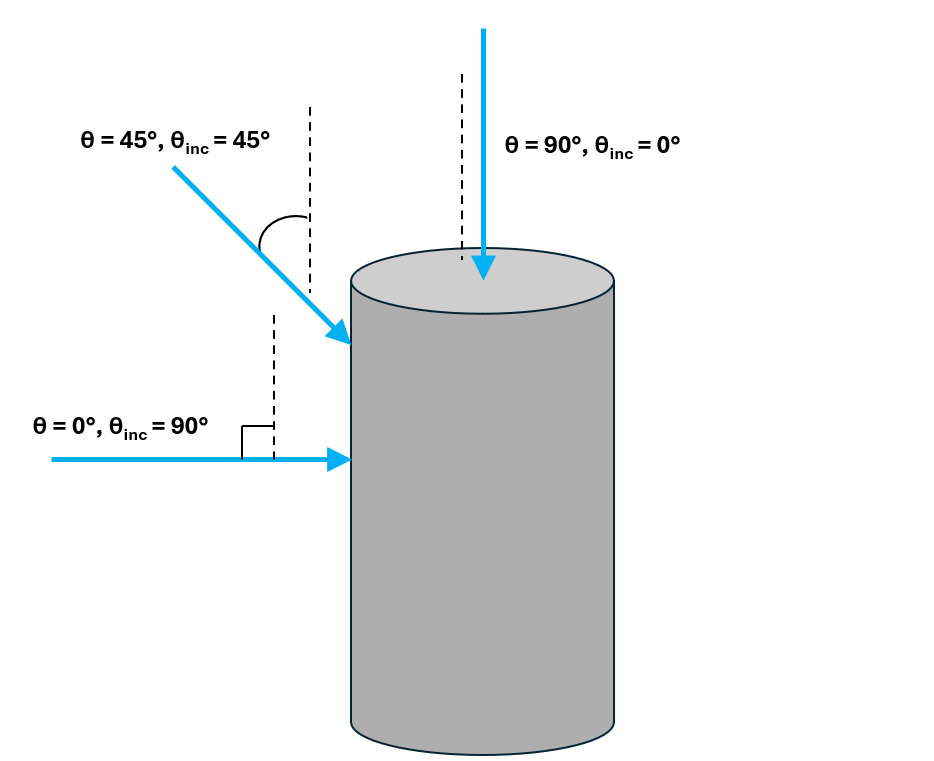}
    \caption{Relationship between the laboratory rotation angle $\theta$ and the neutron
    incidence angle $\theta_{\rm inc}$ relative to the cylinder axis.}
    \label{fig:iso-geometry}
\end{figure}

The laboratory rotation angle $\theta$ and the incidence angle $\theta_{\rm inc}$ are
complementary, as illustrated in \cref{fig:iso-geometry}. The side-on orientation
($\theta=\SI{0}{\degree}$) corresponds to incidence perpendicular to the cylinder axis
($\theta_{\rm inc}=\SI{90}{\degree}$), while the face-on orientation corresponds to axial
incidence. Thus,
\begin{equation}
    \theta_{\rm inc}=\SI{90}{\degree}-\theta
\end{equation}
and $\sin\theta_{\rm inc}=\cos\theta$. This weighting accounts simultaneously for the
directional response of the detector and the distribution of arrival directions in an isotropic
field.

The detector response was simulated at four additional orientations,
\SI{23}{\degree}, \SI{34}{\degree}, \SI{56}{\degree}, and
\SI{68}{\degree}, using a common source-detector geometry. These simulations provide
$R_{\rm sim}(\theta)$, and therefore $w(\theta)$, over the full angular range. The three
measured scaling factors were interpolated onto a fine angular grid, and
Eq.~\ref{eq:siso-continuous} was evaluated as a discrete sum in \SI{5}{\degree} intervals.

Because only three measured scaling factors are available for each configuration, the
interpolation was performed using both piecewise-linear and quadratic functions. The mean of the
two resulting values of $S_{\rm iso}$ was adopted as the central scaling factor. Rather than
using half their separation, the full absolute difference between the two interpolation results
was taken as a conservative systematic uncertainty. The statistical uncertainty propagates
the Poisson uncertainties in the measured and simulated count rates, while the total systematic
uncertainty combines the interpolation uncertainty with that arising from the simulated
geometry.

The resulting isotropic scaling factors are listed in
\cref{tab:iso-scaling-factors}. These uncertainties are propagated into the underground
unfolding by scaling the response functions according to the central, statistical, and systematic
variations of $S_{\rm iso}$ and unfolding the resulting response matrices separately.

\begin{table}[ht]
\caption{Response-weighted isotropic scaling factors for the seven BCS configurations.}
\label{tab:iso-scaling-factors}
\renewcommand{\arraystretch}{1.35}
\bgroup
\setlength{\tabcolsep}{5pt}
\begin{tabular}{lc}
Configuration & $S_{\rm iso}$ \\
\colrule
Bare & 1.000 $\pm$ 0.238$_{\rm stat.}$ $\pm$ 0.055$_{\rm sys.}$ \\
C1 & 0.897 $\pm$ 0.033$_{\rm stat.}$ $\pm$ 0.020$_{\rm sys.}$ \\
C2 & 1.028 $\pm$ 0.010$_{\rm stat.}$ $\pm$ 0.029$_{\rm sys.}$ \\
C3 & 1.147 $\pm$ 0.006$_{\rm stat.}$ $\pm$ 0.024$_{\rm sys.}$ \\
C4 & 1.019 $\pm$ 0.004$_{\rm stat.}$ $\pm$ 0.023$_{\rm sys.}$ \\
C5 & 0.917 $\pm$ 0.004$_{\rm stat.}$ $\pm$ 0.032$_{\rm sys.}$ \\
C6 & 0.835 $\pm$ 0.003$_{\rm stat.}$ $\pm$ 0.021$_{\rm sys.}$ \\
\end{tabular}
\egroup
\end{table}

\begin{figure}[ht]
    \centering
    \includegraphics[width=\linewidth]
    {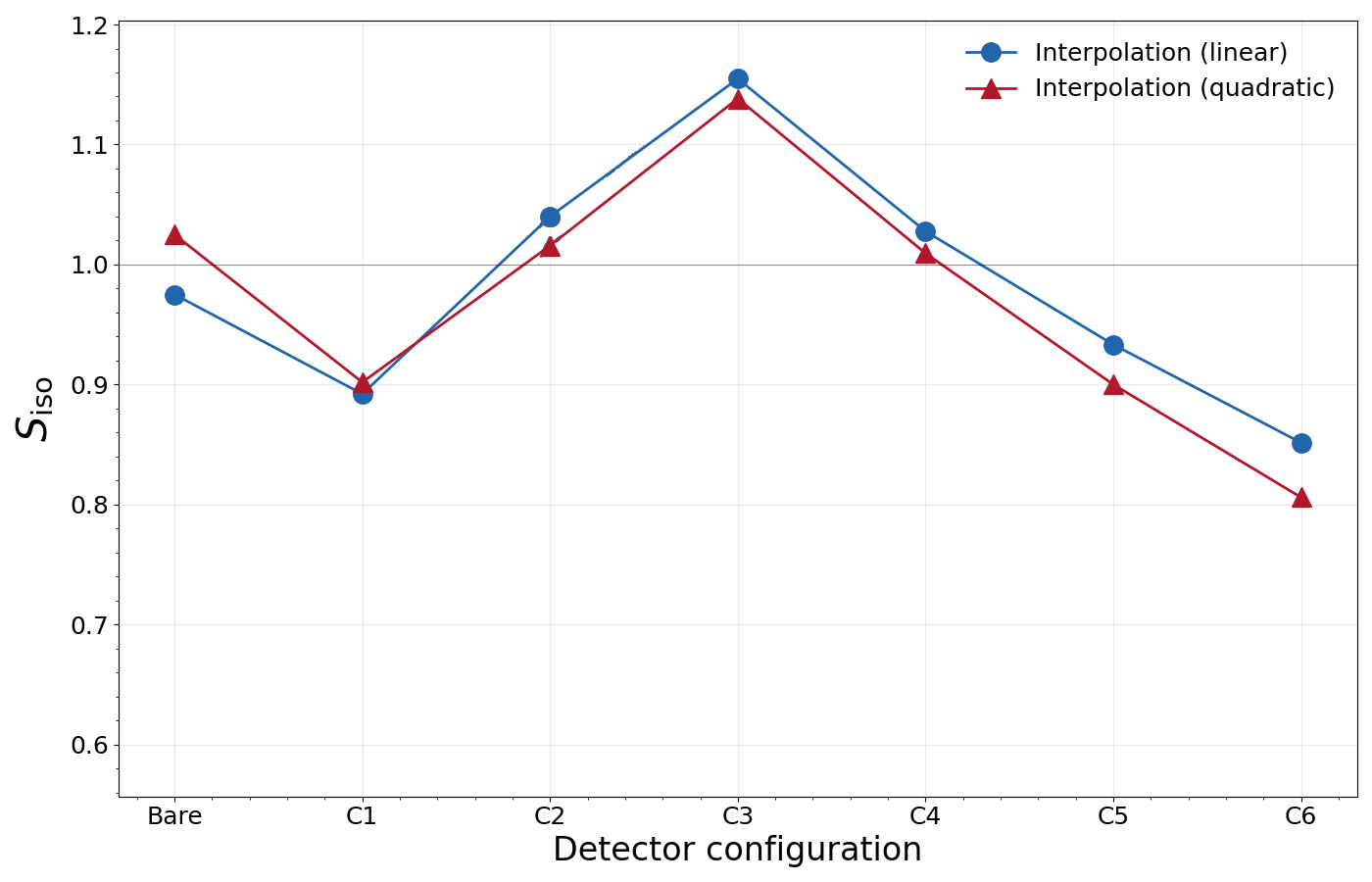}
    
    \caption{Response-weighted isotropic scaling factors obtained using piecewise-linear and quadratic interpolation of the three measured angular scaling factors for each BCS configuration.}
    
    \label{fig:all-scaling-factors}
\end{figure}

%\cref{fig:all-scaling-factors} compares the isotropic scaling factors obtained using a three-point sum with those obtained from the linear and quadratic interpolations. The two
%interpolated results agree closely for all configurations, indicating that the refined scaling
%factors are not strongly dependent on the interpolation method. The three-point estimate deviates
%most noticeably for C1 because its coarse angular average is strongly influenced by the
%anomalously low side-on scaling factor. Sampling the angular response more finely reduces this
%sensitivity, giving a refined value that is consistent with the overall
%behavior of the remaining configurations.
\cref{fig:all-scaling-factors} compares the isotropic scaling factors
obtained using piecewise-linear and quadratic interpolation of the
three measured angular scaling factors. The two methods give similar
results across all seven configurations. Their full absolute difference
is included in the systematic uncertainty to account for sensitivity
to the interpolation choice.

\subsection{Underground Measurements}\label{sec:underground-measurements}

The BCS measurements were performed in the Subatomic Particle Hideout
with the detector deployed near the south wall, approximately
\SI{0.5}{\m} above the floor and at least \SI{0.5}{\m} from each wall.
The $^{3}$He counter was oriented vertically, and the detector assembly
was kept at the same position throughout the campaign. The unfolding
uses the isotropic response functions, assuming an approximately
isotropic local neutron field. Each configuration was measured in a
single continuous live-time window, with run lengths chosen to bring
the statistical uncertainty on the integrated count rate below
approximately \SI{5}{\percent}.

The channel-to-energy mapping was fixed by fitting the full-energy thermal-capture peak in a
high-statistics PuBe dataset and anchoring it to the known \SI{764}{\keV} $Q$-value. The neutron
ROI discussed earlier is applied consistently to every measurement. Its lower bound
corresponds to the \SI{191}{\keV} triton edge, with an uncertainty of $\pm 5$ channels reflecting
the ambiguity in locating the edge. This uncertainty is propagated as a systematic uncertainty on
the extracted counts. The upper bound lies several channels above the full-energy peak to ensure
its complete inclusion. The resulting ROI retains the $^{3}$He$(n,p)^{3}$H events while rejecting
the lower-energy region dominated by ambient $\gamma$ rays and electronic noise.

\begin{figure}[!htb]
    \centering
    \includegraphics[width=\linewidth]{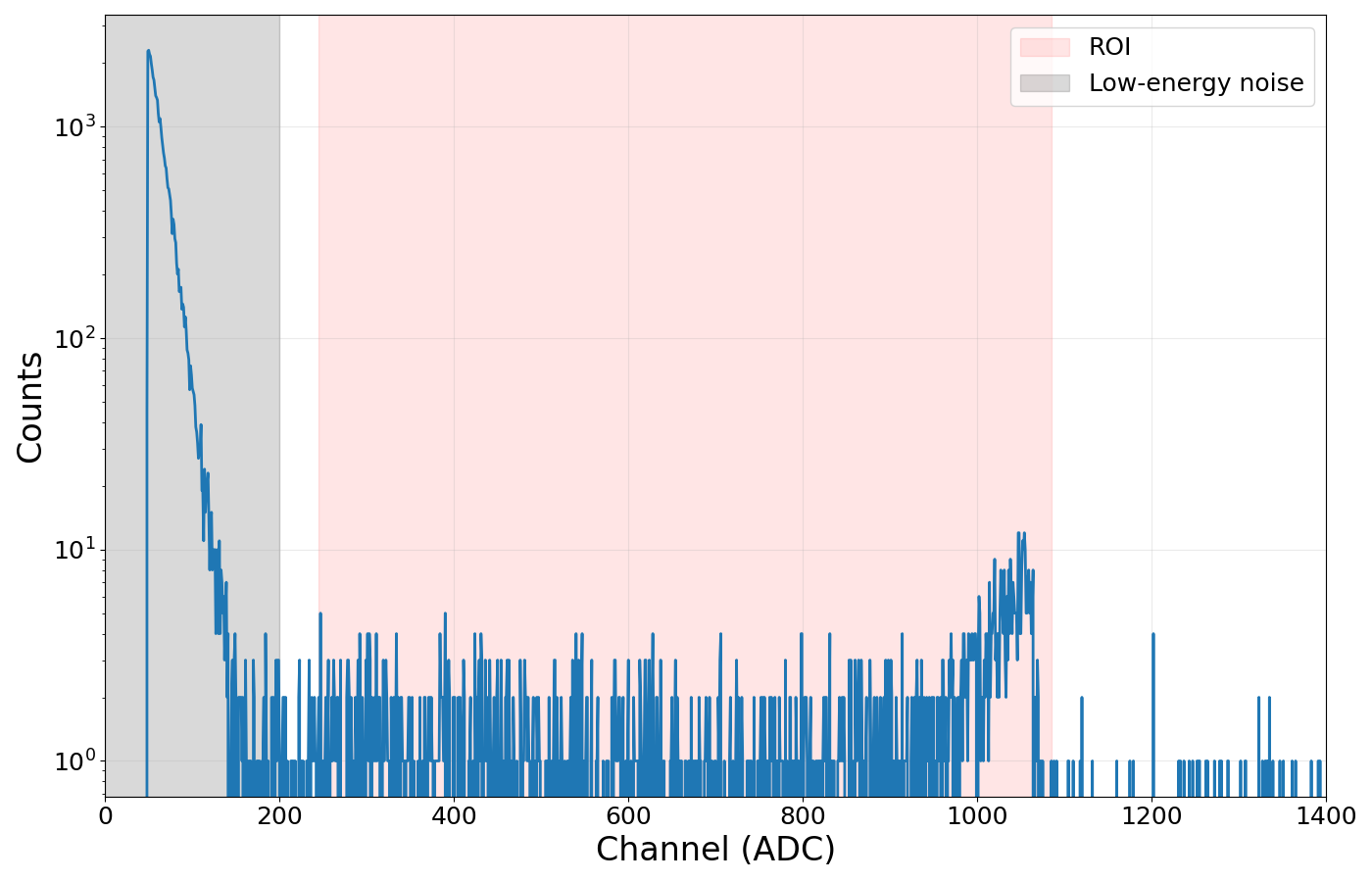}
    
    \caption{A representative underground spectrum (C1 configuration) acquired
    with the Faraday cage in place, with the low-energy noise region (gray) and the
    thermal-capture ROI spanning channels 245--1085 (red), indicated. The residual noise lies entirely
    below the ROI, so it does not contribute to the extracted neutron counts.}
    
    \label{fig:roi-noise}
\end{figure}

During the campaign an unexpected source of electronic noise appeared in the low-energy region of
the pulse-height spectrum. To suppress it, the $^{3}$He tube and charge-sensitive preamplifier
were enclosed in a Faraday cage. The noise was confined to channels below $\approx 200$, well
separated from the ROI whose lower edge lies at channel 245, and because the noise population lies
entirely below the ROI it makes a negligible contribution to the integrated capture counts
(\cref{fig:roi-noise}). The extracted count rates are therefore insensitive to any residual noise
not removed by the cage. The measured data are summarized in \cref{tab:bcs-sph-data}, where the
statistical errors are from counting and the systematic errors are dominated by varying the lower ROI boundary by $\pm~5$ channels.

\begin{table}[ht]
    \caption{Summary of the experimental data taken at the Subatomic Particle Hideout.}
    \label{tab:bcs-sph-data}
%\begin{ruledtabular}
    \renewcommand{\arraystretch}{1.35}
    \bgroup % Keeps the change local to this group
    \setlength{\tabcolsep}{3pt} % Increase padding (Default is 6pt)
\begin{tabular}{l c c c}
Name & Counts & $T_{\rm live}$ (\si{\hour}) &  $R_{\rm exp}$ (cps)  \\ \hline
Bare & 823  & 164.44 & (1.390 $\pm$ 0.049$_{\rm stat.}$ $\pm$ 0.012$_{\rm sys.}$)$\times$10$^{-3}$ \\
C1   & 1162 & 242.29 & (1.332 $\pm$ 0.039$_{\rm stat.}$ $\pm$ 0.010$_{\rm sys.}$)$\times$10$^{-3}$ \\
C2   & 957  & 159.22 & (1.670 $\pm$ 0.054$_{\rm stat.}$ $\pm$ 0.011$_{\rm sys.}$)$\times$10$^{-3}$ \\
C3   & 740  & 165.21 & (1.244 $\pm$ 0.045$_{\rm stat.}$ $\pm$ 0.017$_{\rm sys.}$)$\times$10$^{-3}$ \\
C4   & 868  & 264.22 & (9.125 $\pm$ 0.315$_{\rm stat.}$ $\pm$ 0.063$_{\rm sys.}$)$\times$10$^{-4}$ \\
C5   & 619  & 310.60 & (5.536 $\pm$ 0.224$_{\rm stat.}$ $\pm$ 0.045$_{\rm sys.}$)$\times$10$^{-4}$ \\
C6   & 693  & 498.32 & (3.863 $\pm$ 0.150$_{\rm stat.}$ $\pm$ 0.061$_{\rm sys.}$)$\times$10$^{-4}$ \\
\end{tabular}
\egroup
%\end{ruledtabular}
\end{table}

A direct comparison between the underground count-rate distribution (\cref{tab:bcs-sph-data}) and
that measured with the PuBe source (\cref{tab:bcs-calibration-summary}) already reveals the
qualitative character of the underground field, before any unfolding is performed. For the PuBe
source, whose mean neutron energy is several \si{\MeV}, the response peaks at the
intermediate-to-large moderators, reaching a maximum at C4 and falling off sharply for the bare and
thin configurations that cannot thermalize such a hard spectrum. The underground distribution is
markedly different: the response is largest for the bare and thin moderators (Bare--C2) and
decreases monotonically toward the largest configurations. This shift of the response peak toward
thinner moderators is the signature of a much softer incident spectrum, in which a substantial
fraction of neutrons are already thermal or epithermal and are detected efficiently with little or
no added moderation, while thick moderators serve only to increase absorption. This is precisely
the behavior expected of a radiogenic field dominated by $(\alpha,n)$ and spontaneous-fission
neutrons that have been moderated through scattering in the surrounding rock, and it anticipates
the thermal- and epithermal-dominated spectrum recovered by the unfolding below.

% FIG: roi-noise-plot.png (with noise photo), bcs-faraday.jpg [optional]
% TABLE: tab:bcs-sph-data

\subsubsection{Input Spectrum}

Unlike the PuBe characterization, for which the source spectrum is well known,
the underground neutron spectrum is itself the quantity being measured.
Its reconstruction therefore requires an assumed input spectrum.
Radiogenic neutron production can be estimated from the elemental composition
and uranium and thorium activities of the surrounding
rock~\cite{mei-alpha-n,alpha-n-depth,PARVU2025112035}.
Although the major-element composition of the host rock was characterized
by scanning electron microscopy for the simulations in Ref.~\cite{dakota2},
the uranium and thorium activities in the rock immediately surrounding
the SPH have not been measured. We therefore adopt a physically motivated
default spectrum.

\begin{figure}[ht]
    \centering
    \includegraphics[width=\linewidth]{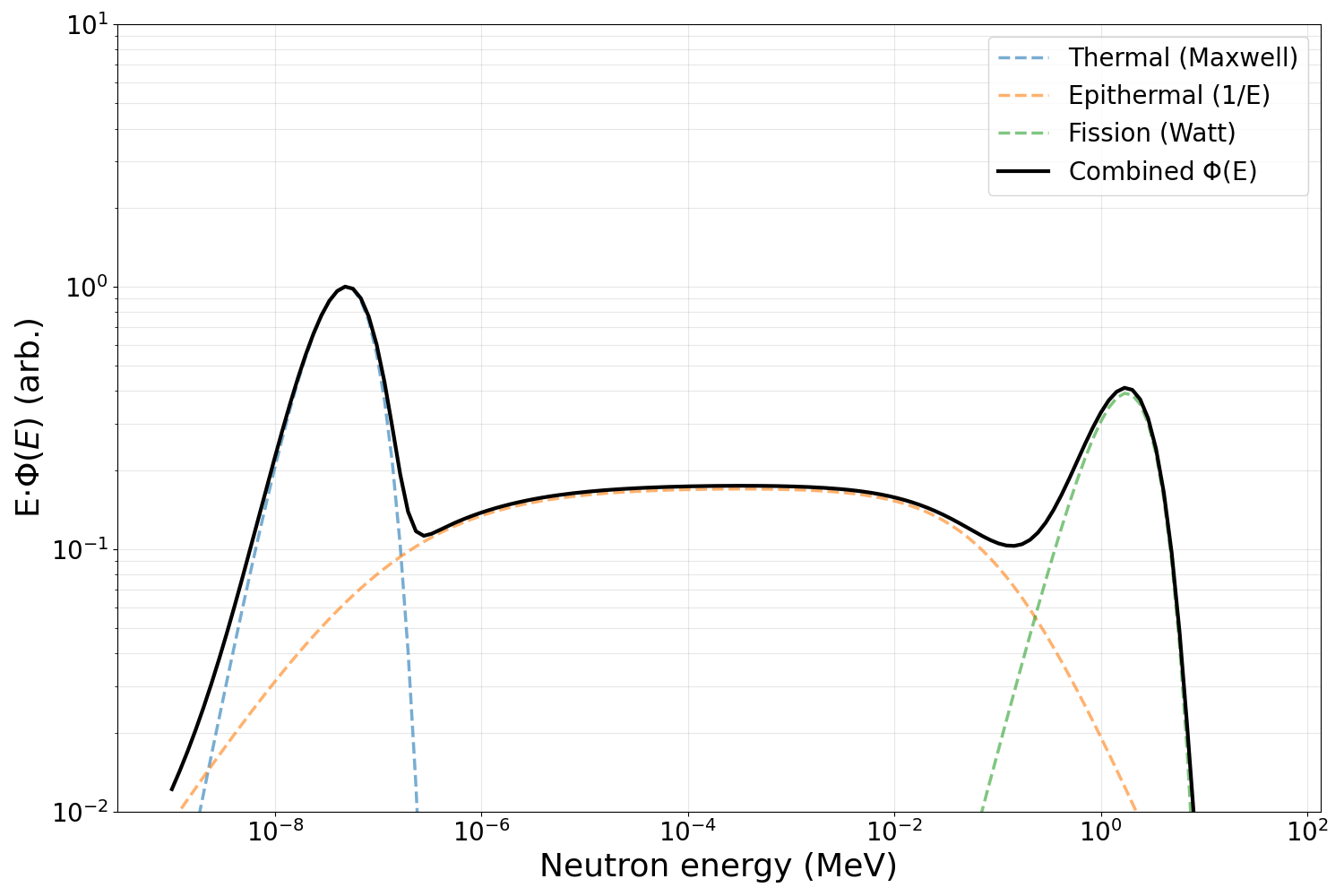}
    \caption{Example default spectrum used for the underground unfolding, constructed from a
    Maxwellian thermal component, a $1/E$ epithermal component, and a Watt spontaneous-fission
    component for $^{238}$U. The spectrum is shown as $E\,\Phi(E)$ in arbitrary units because only
    its relative shape is supplied to the unfolding algorithms.}
    \label{fig:radiogenic-neutron-spectrum}
\end{figure}

The default spectrum contains three components: a Maxwellian distribution for thermal neutrons,
a $1/E$ distribution for the epithermal region produced by repeated downscattering, and a Watt
spectrum describing neutrons from the spontaneous fission of $^{238}$U in the surrounding
rock~\cite{watt-spectrum,mei_hime,lngs-alpha-n-model}. An example is shown in \cref{fig:radiogenic-neutron-spectrum}, with the three components identified separately. Only the
relative spectral shape is supplied to the unfolding algorithms; the absolute normalization is
determined by the measured count rates. The input and response functions use the same 140
logarithmically spaced energy bins.

%Because the underground spectrum is not known beforehand, its reconstruction may depend on the
%assumed input. This dependence was tested using two deliberately dissimilar default spectra, one
%dominated by the thermal component and the other by the fast component. Each spectrum was unfolded
%using \textsc{maxed}, \textsc{gravel}, and \textsc{mlem}, yielding six reconstructions.
%\textsc{maxed} tends to preserve the default spectrum where the measurements provide limited
%information, whereas \textsc{gravel} and \textsc{mlem} iteratively modify it to reproduce the
%measured count rates. The spread among the six reconstructed spectra therefore quantifies the
%residual dependence on both the unfolding algorithm and the assumed input and is included in the
%systematic uncertainty on $\Phi_n$.
Dependence on the assumed input was tested using two default spectra
constructed from the same Maxwellian, $1/E$, and Watt components,
but with substantially different relative weights: one dominated by
the thermal component and the other by the fast component.
The measured count rates were unfolded with each default spectrum
using \textsc{maxed}, \textsc{gravel}, and \textsc{mlem}, yielding
six reconstructions. Their spread is included in the systematic
uncertainty on $\Phi_n$ as an estimate of sensitivity to the selected
algorithms and component weightings.

% FIG: radiogenic-neutron-spectrum.png

\subsection{Results}\label{sec:underground-results}

\begin{figure}[ht]
    \centering
    \includegraphics[width=\linewidth]{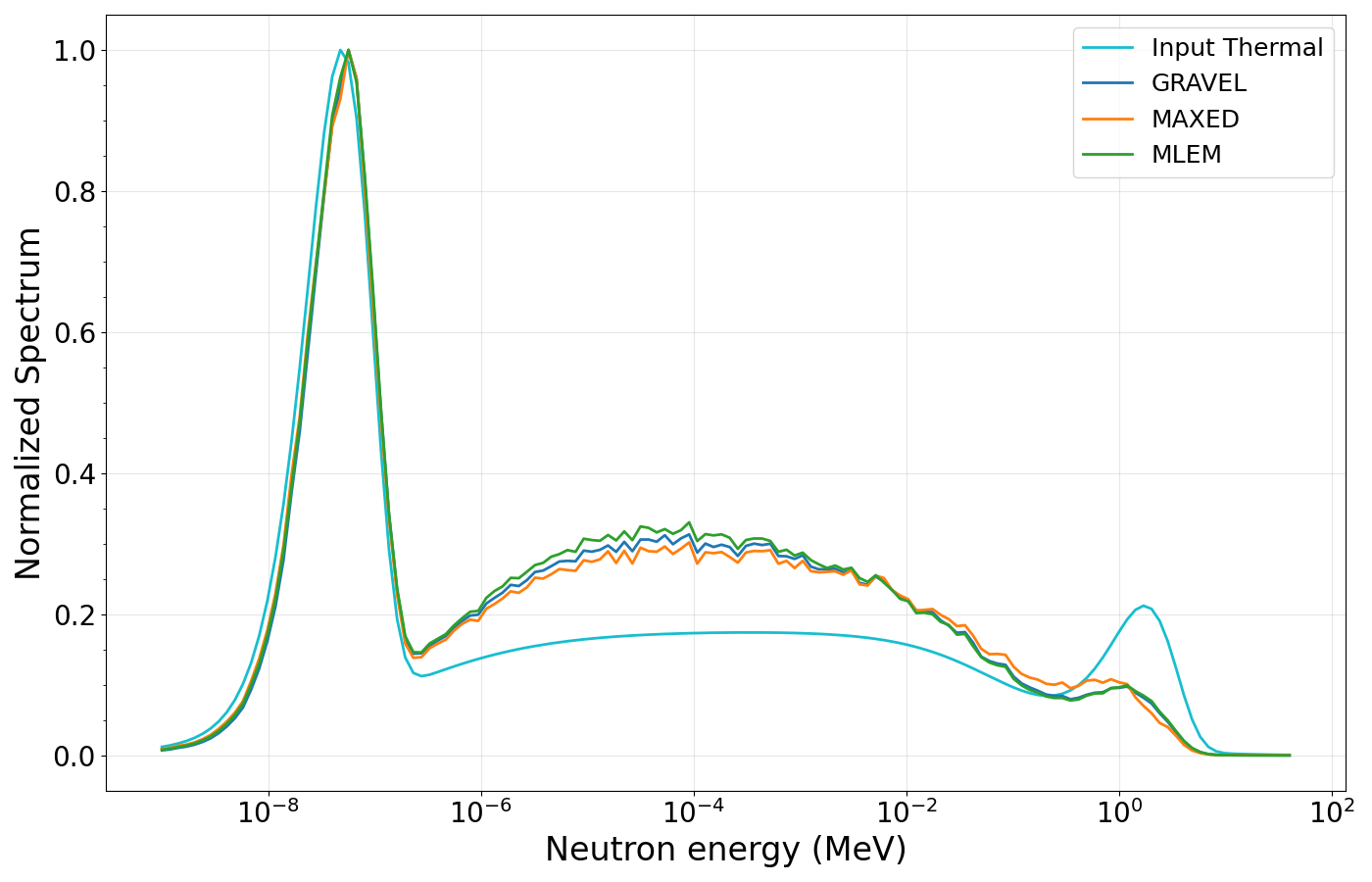}
    \\[4pt]
    \includegraphics[width=\linewidth]{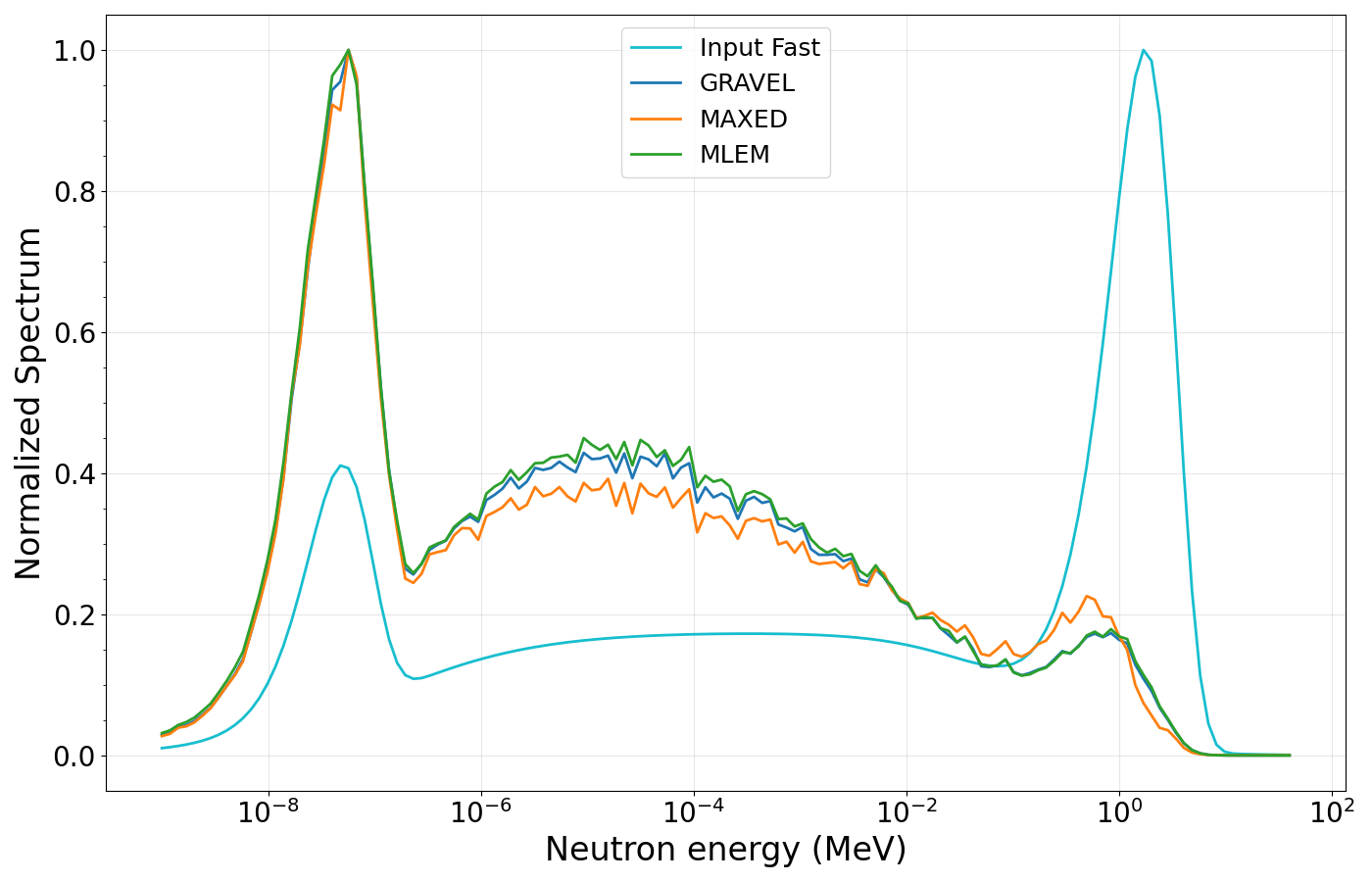}
    \caption{Unfolding results from \textsc{gravel}, \textsc{maxed}, and \textsc{mlem} for the two
    input spectra. \emph{Top:} thermal-dominated input. \emph{Bottom:} fast-dominated input. The
    corresponding input is overlaid in each panel.}
    \label{fig:unfolding-comparisons}
\end{figure}

\cref{fig:unfolding-comparisons} compares the six reconstructions,
grouped by default spectrum with the corresponding input overlaid.
The reconstructed spectra exhibit consistent broad thermal,
epithermal, and fast components despite the different input
weightings. This agreement indicates stability within the tested
set of algorithms and default spectra. It does not independently
establish the detailed shapes of the components, which remain
influenced by the input spectra where the response functions
provide limited information.

\begin{figure}[ht]
    \centering
    \includegraphics[width=\linewidth]{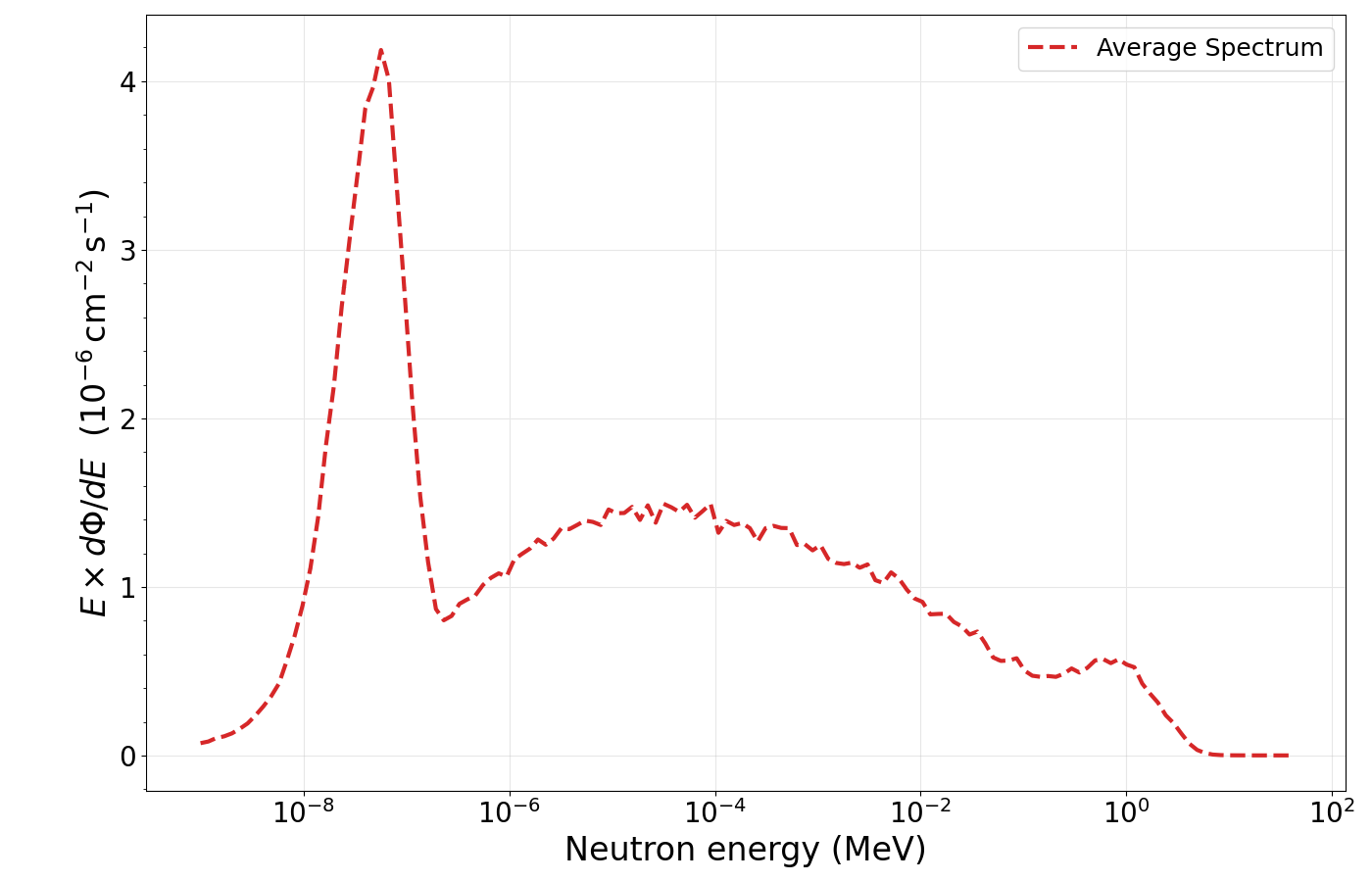}
    \caption{Unfolded local scalar neutron spectrum at the Subatomic Particle Hideout,
dominated by radiogenic production and taken as the average of the six output spectra.}
    \label{fig:sph-unfolded-average-spectrum}
\end{figure}

The final spectrum (\cref{fig:sph-unfolded-average-spectrum}) is
taken as the arithmetic mean of the six reconstructions.
Integrating this spectrum gives a total local scalar neutron flux at the detector for
$E_{\rm n}\lesssim\SI{10}{\MeV}$ of
\begin{equation}
    \Phi_{\rm n}
    = (2.51 \pm 0.58_{\rm stat.} \pm 0.39_{\rm sys.})
    \times10^{-5}~
    \si{\per\square\centi\meter\per\second}\,.
\end{equation}

The integrated-flux uncertainty includes contributions from the
measured counts, response-function scaling factors, and unfolding
procedure. The counting contribution is evaluated by varying the
input count data by the uncertainties in \cref{tab:bcs-sph-data}.
For the scaling contribution, the response functions are scaled
using the central $S_{\rm iso}$ values and their upper and lower
statistical and systematic bounds. The five resulting response
matrices are unfolded separately, and deviations from the nominal
result determine the propagated uncertainties. The statistical
part originates in the Poisson uncertainties of the measured and
simulated calibration counts; the systematic part combines the
interpolation spread and simulated-geometry uncertainty
(Sec.~\ref{sec:isotropic-corrections}). The unfolding contribution
is systematic and is assigned as the maximum deviation of the six
reconstructions from their mean. Statistical and systematic
contributions are summed separately in quadrature.

\begin{figure}[!htb]
    \centering
    \includegraphics[width=\linewidth]{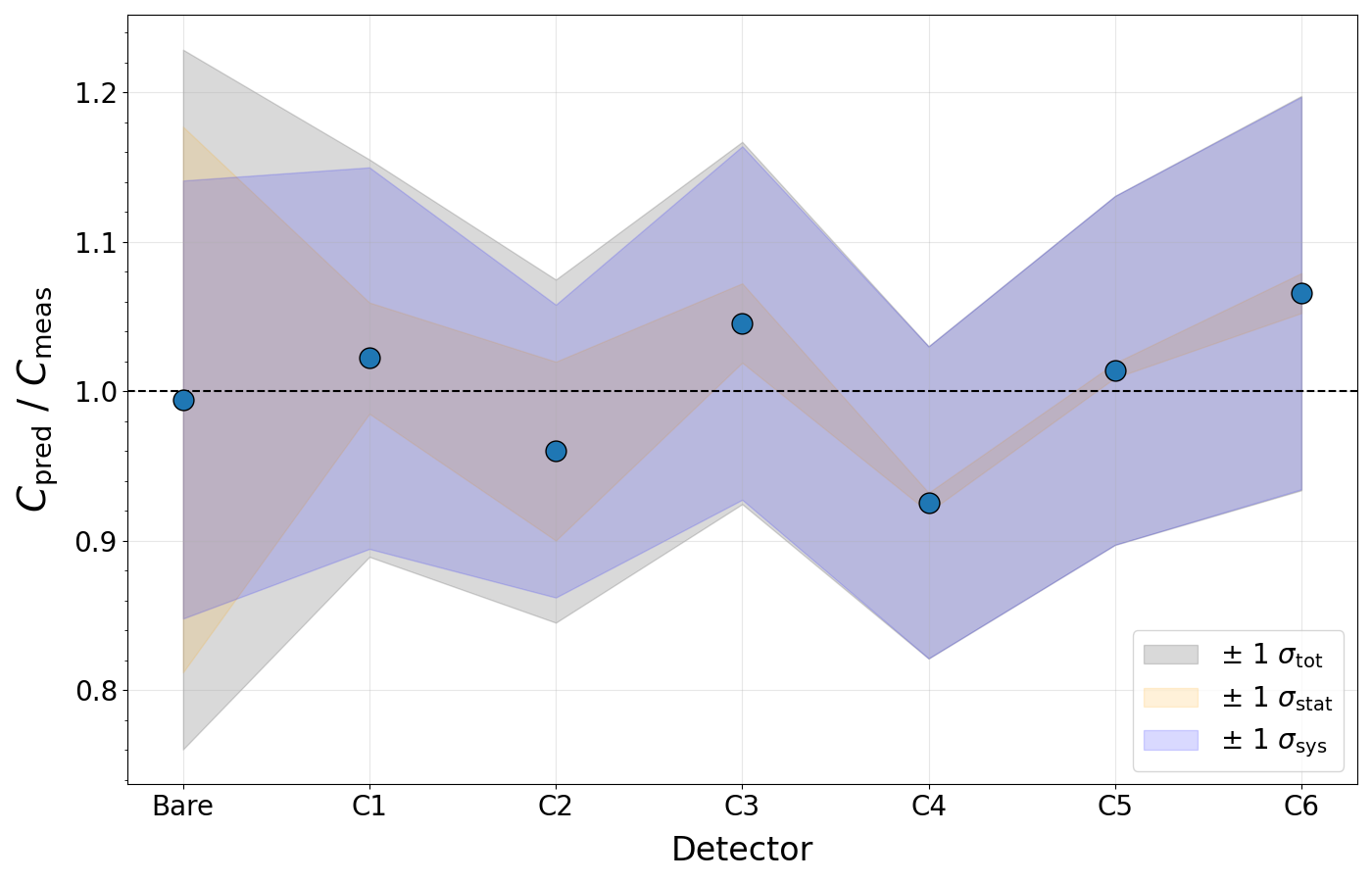}
    \caption{Ratio of predicted to measured counts for each detector configuration, obtained by
    forward-folding the averaged output spectrum with the response functions. The total error is
    the quadrature sum of statistical and systematic contributions.}
    \label{fig:sph-predicted-count-ratios}
\end{figure}

As a final check, the averaged spectrum was forward-folded with the response functions to predict
the count rate for each configuration. The ratio $C_{\rm pred}/C_{\rm meas}$, shown in
\cref{fig:sph-predicted-count-ratios}, quantifies the agreement between the reconstructed and
measured rates. All seven configurations are consistent with the measurements, with a maximum
relative difference of approximately \SI{8}{\percent}. The bare configuration has the largest
uncertainty because of its low count rate during the PuBe characterization and the resulting
statistical uncertainty in its response-function scaling factor. This calibration uncertainty,
rather than the counting statistics of the underground measurement, is the dominant statistical
limitation associated with the bare response function and propagates into the uncertainty on the
integrated flux.

The integrated flux was also evaluated over three energy intervals: thermal
($E_{\rm n}\leq\SI{0.5}{\electronvolt}$), epithermal
($\SI{0.5}{\electronvolt}<E_{\rm n}\leq\SI{0.1}{\MeV}$), and fast
($E_{\rm n}>\SI{0.1}{\MeV}$). The contribution reconstructed above \SI{10}{\MeV} is negligible
relative to the total flux, so no upper bound is imposed on the fast interval. Because the BCS
sensitivity decreases at these energies, no detailed interpretation of the spectrum above
\SI{10}{\MeV} is attempted.

\cref{sph-radiogenic-flux} compares the resulting integrated fluxes with measurements from other
underground facilities. The CURIE values are consistent with previous measurements, particularly
those from facilities with similar rock compositions~\cite{YOON2021102533,kamioka_3he}. At CURIE,
the epithermal component accounts for approximately \SI{57}{\percent} of the total flux, consistent
with radiogenic neutrons undergoing repeated scattering in the surrounding rock and experimental
hall before reaching the detector. The order-of-magnitude variation among facilities with
overburdens ranging from \SI{140}{} to \SI{6230}{}~m.w.e.\ further demonstrates that radiogenic
neutron production depends more strongly on the local uranium and thorium content and host-rock
composition than on facility depth.

\begin{table*}[ht]
\caption{Comparison of measured radiogenic neutron fluxes at underground facilities. Three energy
ranges are shown where available. Uncertainties are the quadrature-summed total of statistical and
systematic contributions.}
\label{sph-radiogenic-flux}
%\begin{ruledtabular}
\renewcommand{\arraystretch}{1.35}
%\bgroup % Keeps the change local to this group
\setlength{\tabcolsep}{8pt} % Increase padding (Default is 6pt)
\begin{threeparttable}
\begin{tabular}{l c c c c c}
Underground Facility & Depth & Thermal & Epithermal & Fast  & Total \\
 & (m.w.e.) & $E_{n} \leq \SI{0.5}{\electronvolt}$ & \SI{0.5}{\electronvolt} $< E_{n} \leq$ \SI{0.1}{\MeV} & \SI{0.1}{\MeV} $< E_{n}$ & (\si{\per\square\centi\meter\per\second})$\times10^{-6}$ \\ \hline
Felsenkeller~\cite{felsenkeller-alpha-n}\tnote{a} & 140 & --- & --- & --- & 235 $\pm$ 43\\
CURIE (this work) & 406 & 9.09 $\pm$ 1.55 & 14.40 $\pm$ 4.63 & 1.63 $\pm$ 0.87 & 25.10 $\pm$ 6.99\\
JUSL~\cite{alpha-n-depth} & 1604 & 34.0 $\pm$ 1.3 & 27.2 $\pm$ 0.6 & 47.0 $\pm$ 1.6 & 107.0 $\pm$ 7.0\\
Y2L~\cite{YOON2021102533} & 2000 & 14.4 $\pm$ 1.5 & --- & --- & 44.6 $\pm$ 6.6\\
Soudan~\cite{surf-soudan-alpha-n} & 2032 & 0.7 $\pm$ 0.16 & --- & --- & ---\\
Yemilab~\cite{yemilab-alpha-n}\tnote{a} & 2350 & 14.1 $\pm$ 0.9 & --- & 3.1 $\pm$ 0.7 & 35.0 $\pm$ 1.7\\
Canfranc (Hall A)~\cite{canfranc-alpha-n} & 2450 & 3.4 $\pm$ 0.2 & --- & --- & 14.8 $\pm$ 0.2\\
Kamioka~\cite{kamioka_3he} & 2700 & 7.88 & --- & --- & 23.5 $\pm$ 0.7$^{+1.9}_{-2.1}$\\
LNGS~\cite{lngs-alpha-n} & 3100 & 2.05 $\pm$ 0.06 & 1.28 $\pm$ 0.31 & 2.56 $\pm$ 0.27 & 5.89 $\pm$ 0.64\\
SURF~\cite{surf-soudan-alpha-n} & 3650 & 9.9 $\pm$ 1.2 & --- & --- & ---\\
Modane~\cite{modane-alpha-n} & 4080 & 3.57 $\pm$ 0.32 & --- & --- & 9.6\\
SURF~\cite{surf-soudan-alpha-n} & 4230 & 1.7 $\pm$ 0.3 & --- & --- & ---\\
CJPL~\cite{cjpl-alpha-n} & 6230 & 7.03 $\pm$ 1.81 & --- & --- & 26.9 $\pm$ 10.2\\
\end{tabular}
%\egroup
\begin{tablenotes}\footnotesize
\item[a] Average of three locations.
\end{tablenotes}
\end{threeparttable}
%\end{ruledtabular}
\end{table*}

\section{Muon-induced Neutrons}\label{sec:muon-neutrons}

The BCS measurement of Sec.~\ref{sec:radio-neutrons} constrains the neutron field up to
approximately \SI{10}{\MeV}, above which the detector sensitivity decreases substantially.
In this higher-energy region, the radiogenic contribution is strongly suppressed and
muon-induced neutrons become the dominant component of the underground neutron background.
Below this energy, separating the two populations would require muon
(anti)coincidence tagging~\cite{zhang_soudan,zhang_mu_n,sutanto_mu_n}, while the muon-induced
contribution remains subdominant to the radiogenic flux~\cite{mei_hime,formaggio_backgrounds}.
The present measurement therefore focuses on the high-energy neutron field, where the cosmogenic
component can be isolated without direct muon tagging.

Muon-induced neutrons were measured using a Bicron BC-501A organic liquid scintillator (OLS) detector~\cite{luxium_bc501a_datasheet}.
The cylindrical active volume has a diameter and height of
\SI{5}{in}, corresponding to approximately \SI{1.61}{\litre}, and is enclosed in a sealed aluminum housing coupled to a \SI{5}{in} photomultiplier tube (PMT). Fast neutrons produce nuclear recoils within
the scintillator, while differences in the temporal response to nuclear and electronic
recoils enable neutron--$\gamma$ discrimination through pulse-shape analysis.
The measured count is converted to an integral neutron flux at the rock--cavern boundary
using the simulated transport and detector response. The reported flux follows the
first-entry, boundary-area convention of Ref.~\cite{dakota2}, allowing a comparison
with the prediction reported there.

\subsection{Fast Neutron Interactions}\label{sec:fast-neutron-interactions}

Fast neutrons deposit energy in an organic liquid scintillator primarily by scattering from
hydrogen and carbon nuclei~\cite{BROWN20181,CHADWICK20141}. In elastic scattering, part of the
neutron kinetic energy is transferred to a recoiling nucleus, which subsequently ionizes the
scintillator. A neutron can transfer up to its full kinetic energy to hydrogen in a single
collision, compared with approximately \SI{28}{\percent} for carbon. Combined with the greater
hydrogen abundance and larger elastic-scattering cross section, this makes proton recoils the
dominant energy-deposition mechanism at neutron energies below a few \si{\MeV}.

At higher energies, inelastic reactions on carbon become increasingly important. These include
excitation of the \SI{4.44}{\MeV} first excited state of $^{12}$C and, above approximately
\SI{8}{\MeV}, breakup reactions such as
$^{12}$C$(n,n')3\alpha$~\cite{n-c-3a,n-c-3a-2}. Energy deposited by the resulting charged
particles, including protons, $\alpha$ particles, and the recoiling nuclear fragments, contributes
to the scintillation signal and must be included when reconstructing the detector response.

\begin{figure}[htb]
    \centering
    \includegraphics[width=\linewidth]{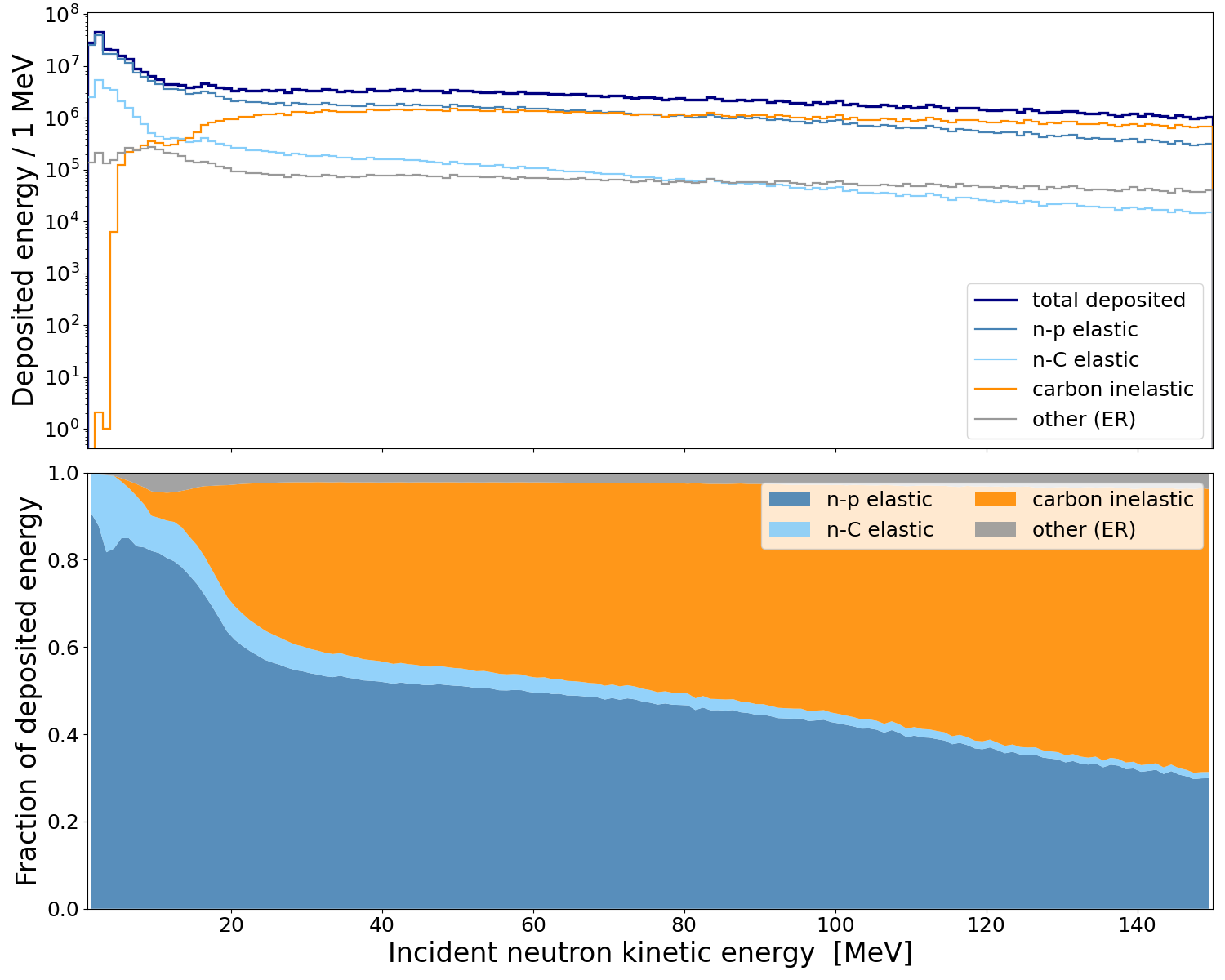}
    \caption{Partition of the energy deposited in BC-501A as a function of incident neutron
    energy, simulated in \textsc{geant4} for $E_n\geq\SI{1}{\MeV}$.
    \emph{Top:} deposited energy per \SI{1}{\MeV} incident-energy bin from $n$--$p$ elastic
    scattering, $n$--C elastic scattering, carbon-inelastic reactions, and electromagnetic
    recoils (ER) produced primarily by de-excitation $\gamma$ rays.
    \emph{Bottom:} fractional contribution of each interaction category to the total deposited
    energy.}
    \label{fig:energy-budget}
\end{figure}

\cref{fig:energy-budget} shows the simulated partition of energy deposited in BC-501A by
neutrons sampled from the muon-induced spectrum predicted for the SPH in Ref.~\cite{dakota2}.
Hydrogen recoils account for approximately \SI{80}{\percent} of the deposited energy below
\SI{10}{\MeV}. The contribution from carbon-inelastic reactions increases above the corresponding
reaction thresholds, becomes comparable to that from $n$--$p$ elastic scattering near
\SIrange{80}{100}{\MeV}, and accounts for approximately two-thirds of the deposited energy at
\SI{150}{\MeV}. When integrated over the incident spectrum above \SI{1}{\MeV}, proton recoils
contribute \SI{62}{\percent} of the deposited energy, while carbon interactions contribute
\SI{36}{\percent}, including \SI{29}{\percent} from inelastic and \SI{7}{\percent} from elastic
scattering. Although proton recoils dominate the spectrum-integrated response, the increasing
contribution from carbon reactions must be included when relating the visible energy to the
incident neutron energy.

\subsection{Pulse Shape Discrimination}\label{sec:psd-procedure}

BC-501A is sensitive to both nuclear and electronic recoils.
In the underground environment, nuclear recoils are produced
predominantly by neutron interactions, while electronic recoils
are induced by $\gamma$ rays and residual cosmogenic muons.
The greater ionization density of proton and carbon recoils
produces a larger slow component in the scintillation light,
giving neutron-induced pulses a longer decay tail than
electronic-recoil pulses of comparable
amplitude~\cite{BIRKS-scintillation,liquid_scint}.
This difference provides the basis for pulse shape
discrimination (PSD).

\begin{figure}[htb]
    \centering
    \includegraphics[width=\linewidth]{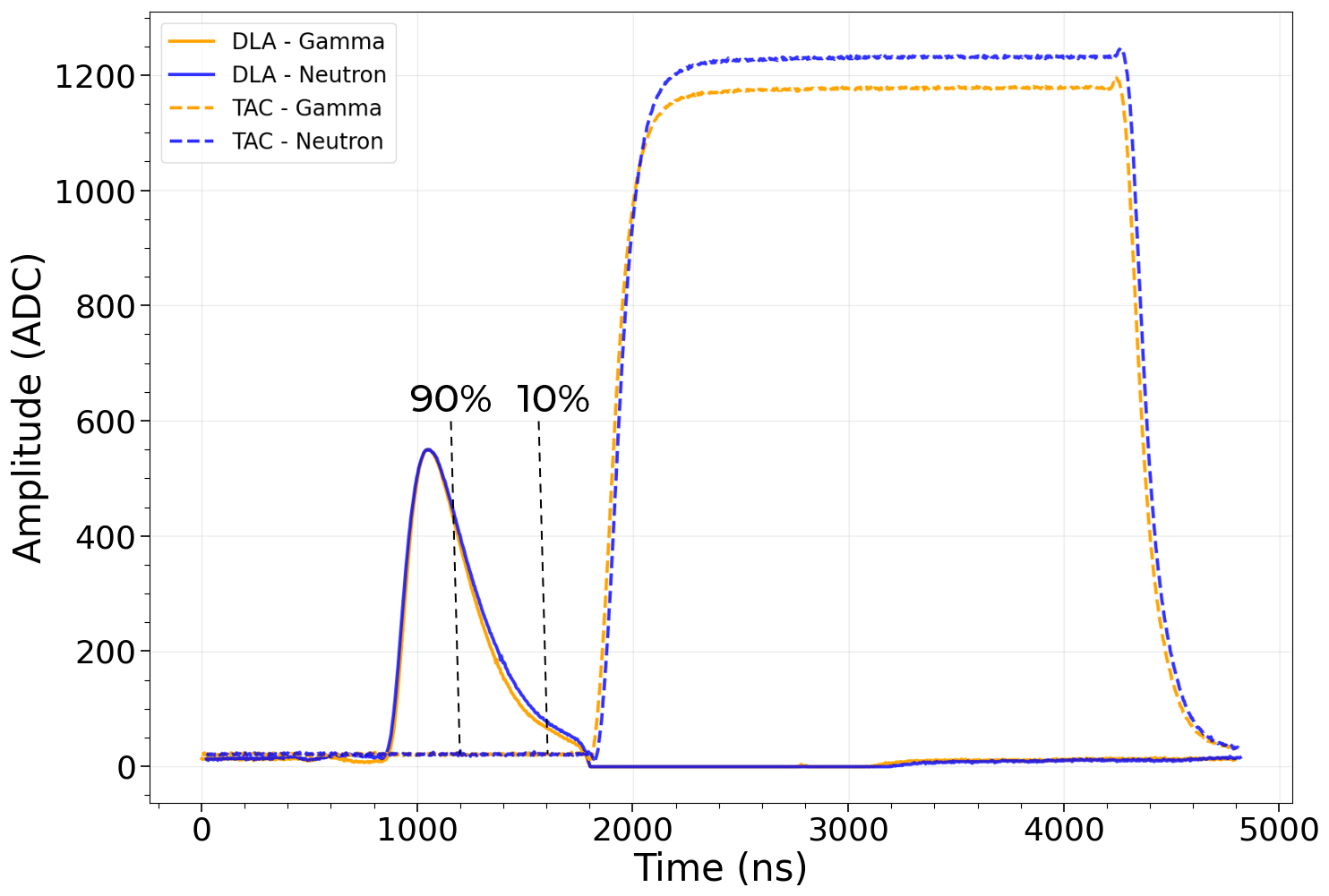}
    \caption{Comparison of the DLA and TAC signals for representative electronic- and
    nuclear-induced pulses of equal DLA amplitude. The longer nuclear-pulse decay time produces a
    larger interval between the \SI{90}{\percent} and \SI{10}{\percent} trailing-edge crossings
    and therefore a larger TAC amplitude.}
    \label{fig:psd-tac-vs-raw-pulse}
\end{figure}

\begin{figure}[htb]
    \centering
    \includegraphics[width=\linewidth]{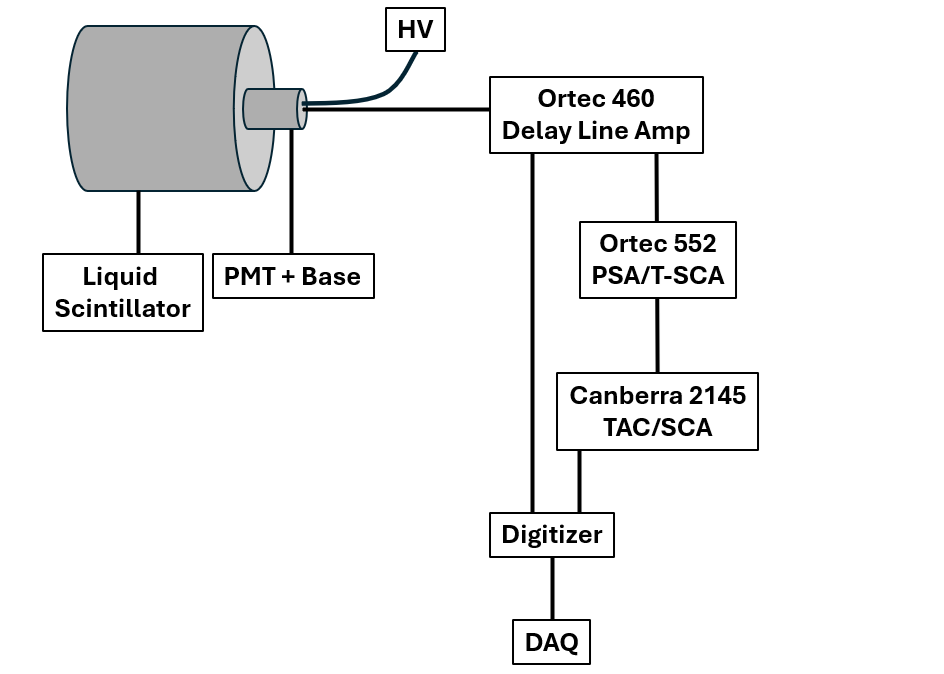}
    \caption{Signal chain used for pulse shape discrimination and waveform acquisition.}
    \label{ls_schematic}
\end{figure}

Several PSD methods are available, including charge integration, zero-crossing, rise-time, and frequency-domain techniques~\cite{psd1,psd2,dft-psd}. We employ an analog decay-time method. The PMT signal is passed through a delay-line amplifier (DLA), whose amplified unipolar output is sent simultaneously to the digitizer and a pulse shape analyzer (PSA). The PSA generates timing signals when the trailing edge crosses \SI{90}{\percent} and \SI{10}{\percent} of the pulse amplitude. These signals provide the start and stop inputs to a time-to-amplitude converter (TAC), whose output amplitude is proportional to the interval between the two crossings. A pulse with a longer decay tail therefore produces a larger TAC amplitude, as illustrated in \cref{fig:psd-tac-vs-raw-pulse}. The complete acquisition chain is shown in \cref{ls_schematic}.

\begin{figure}[htb]
    \centering
    \includegraphics[width=\linewidth]{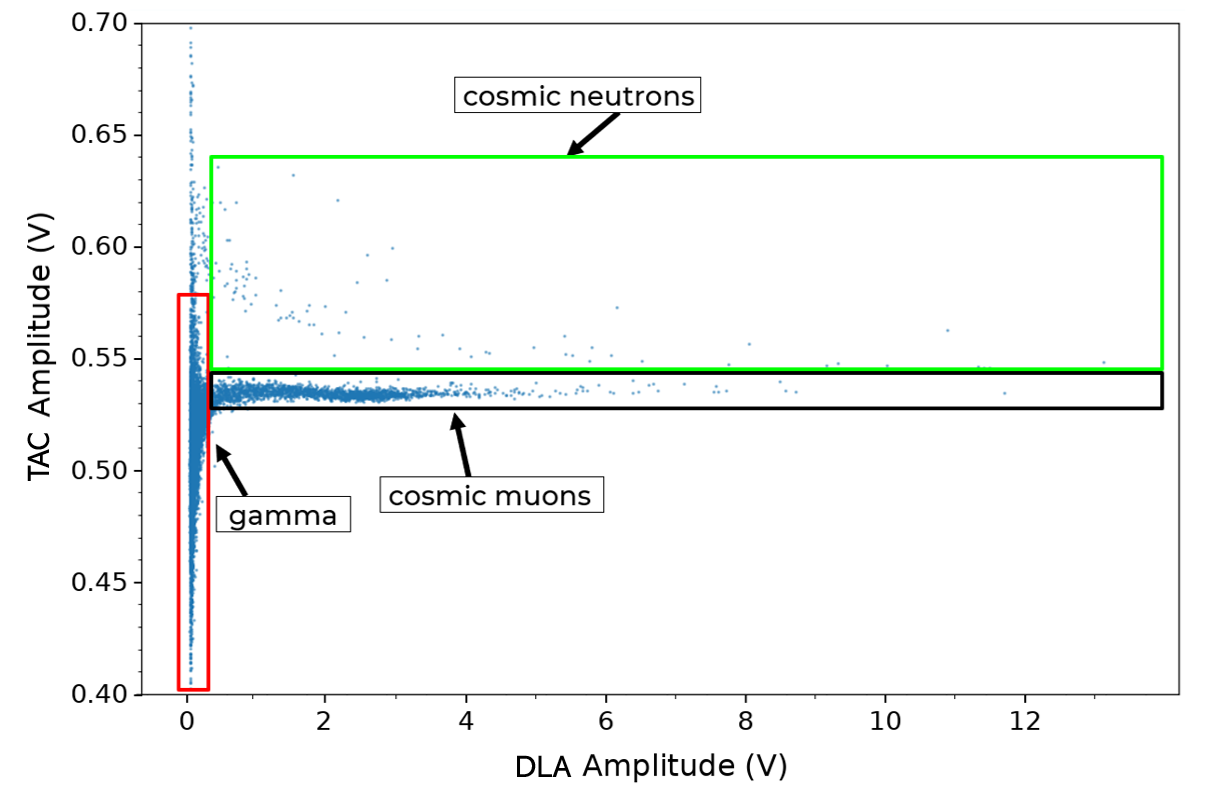}
    \caption{TAC amplitude versus DLA pulse amplitude for the surface-level measurement. The
    upper band contains neutron-like events, while the lower band contains electronic recoils from
    $\gamma$ rays and muons. The highlighted regions identify atmospheric-neutron, muon, and
    low-energy $\gamma$-ray populations used to characterize the discrimination.}
    \label{fig:tac_v_shaped_pulse}
\end{figure}

Plotting the TAC amplitude against the DLA pulse amplitude separates events according to their
decay time. The surface-level data in \cref{fig:tac_v_shaped_pulse} exhibit a neutron-like band
at larger TAC amplitudes and an electronic-recoil band at smaller TAC amplitudes. The populations
are well separated over most of the measured range but begin to overlap at low pulse amplitudes.
In this region, the reduced number of detected photoelectrons increases statistical fluctuations
in the measured pulse shape, while quenching compresses the visible-energy response of proton and
carbon recoils relative to electron recoils~\cite{Birks1965}. The loss of discrimination at low
visible energy ultimately determines the analysis threshold applied to the underground neutron
measurement.

\subsection{Energy Calibration}\label{sec:bc501-calibration}

Unlike the $^{3}$He measurement, which requires only the identification of events within a fixed
ROI, the BC-501A measurement requires the pulse amplitude to be related to the deposited visible
energy. Because Compton scattering dominates the response of the low-$Z$ scintillator to
$\gamma$ rays, standard calibration sources produce continua rather than full-energy
peaks~\cite{gamma-pe,gamma-pp,compton-scatt,compton-scatt2}. The calibration is therefore
anchored to the corresponding Compton edges.

Conventional $\gamma$-ray sources provide calibration points only up to approximately
\SI{3}{\MeV}. The calibration range for this work was extended using the \SI{4.439}{\MeV} de-excitation
$\gamma$ ray from $^{12}$C produced by the PuBe source and the \SI{6.128}{\MeV}
de-excitation $\gamma$ ray from $^{16}$O produced by a $^{244}$Cm-$^{13}$C source. A final
high-energy calibration point was obtained from cosmic-ray muons. Unlike the photon sources,
muons deposit energy directly in the scintillator and produce a minimum-ionizing peak rather than
a Compton continuum. The complete set of calibration energies is listed in
\cref{tab:calibration-sources}.

\begin{table}[htb]
\caption{Calibration sources and associated energies. For the photon sources, the full
$\gamma$-ray energy and corresponding Compton-edge energy are listed. The photon energies are
taken from Ref.~\cite{bnlNuDat}; their uncertainties are much smaller than
\SI{1}{\percent} and are not reported. The muon entry gives the simulated most-probable energy
deposited in the BC-501A detector.}
\label{tab:calibration-sources}
\renewcommand{\arraystretch}{1.35}
\begin{tabular}{lcc}
Source & $E$ (\si{\mega\electronvolt})
       & $E_{\rm CE}$ (\si{\mega\electronvolt}) \\
\colrule
$^{137}$Cs & 0.662 & 0.477 \\
$^{40}$K   & 1.461 & 1.243 \\
$^{208}$Tl & 2.614 & 2.382 \\
$^{12}$C   & 4.439 & 4.197 \\
$^{16}$O   & 6.128 & 5.883 \\
$\mu$      & $21.485 \pm 0.501$ & --- \\
\end{tabular}
\end{table}

\begin{figure}[htb]
    \centering
    \includegraphics[width=\linewidth]{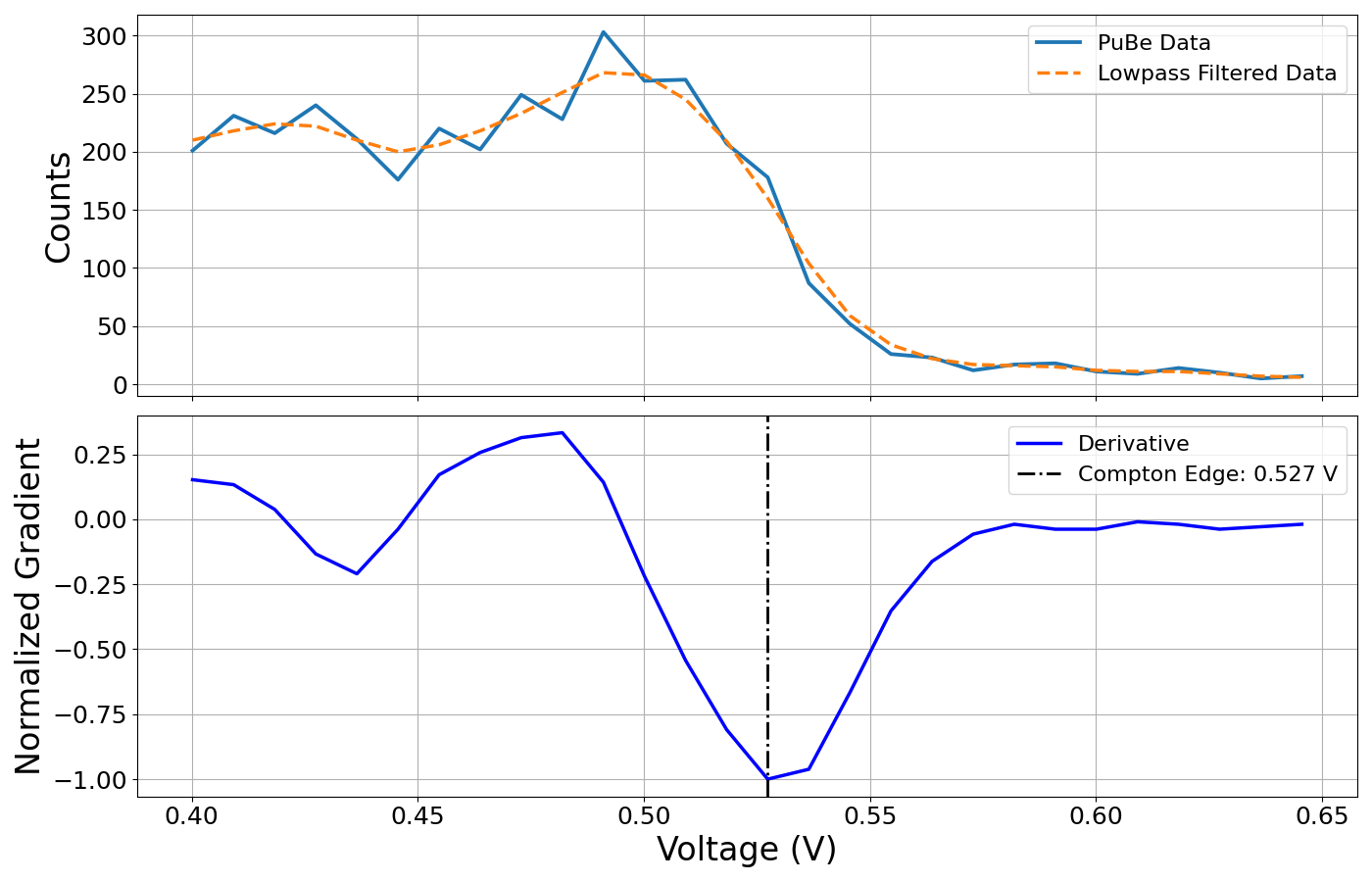}
    \caption{Determination of the PuBe Compton edge. A low-pass filter is applied to the measured
    continuum, and the edge is identified from the minimum of the derivative of the smoothed
    spectrum.}
    \label{pube_calibration}
\end{figure}

Finite detector resolution broadens each Compton continuum and obscures its otherwise sharp endpoint. To determine
the edge consistently, a low-pass smoothing filter was applied to each measured $\gamma$-source spectrum from \cref{tab:calibration-sources} and the
minimum of the derivative of the smoothed distribution was identified~\cite{compton_gradient}. An example for the
\SI{4.439}{\MeV} $\gamma$ ray from the PuBe source is shown in \cref{pube_calibration}.

The scintillation light output is not strictly proportional to deposited energy across the full
dynamic range because the relative contribution of ionization quenching varies with particle
stopping power~\cite{Birks1965,zhang_soudan}. The response below the high-energy muon region was
therefore modeled empirically as a second-order polynomial~\cite{zhang_soudan},
\begin{equation}
    E(V)=aV^2+bV+c,
\end{equation}
constrained by the Compton-edge measurements in \cref{tab:calibration-sources} and the muon
calibration point. The fit was performed using orthogonal-distance regression~\cite{odr} so that
uncertainties in both the measured edge positions and the associated energies were included. Energies are reported in
MeV electron equivalent (MeVee), defined as the energy an electron
would deposit to produce the same detector signal as the interacting
particle.

\begin{figure}[htb]
    \centering
    \includegraphics[width=\linewidth]{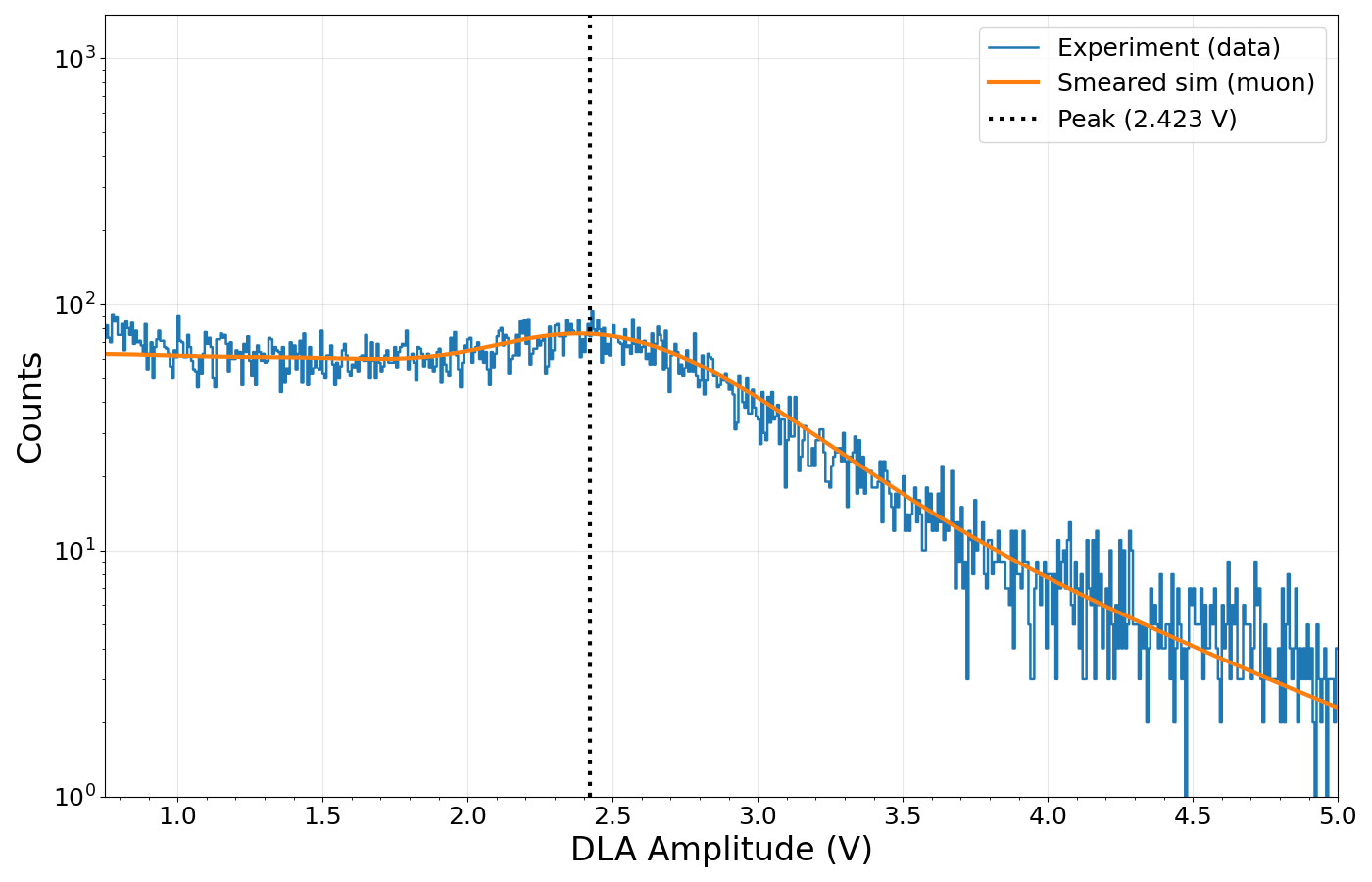}
    \caption{Comparison of the measured muon pulse-height spectrum with the \textsc{geant4}
    energy-deposition simulation after accounting for the finite detector resolution. Agreement
    in the minimum-ionizing peak and high-energy tail constrains the linear high-energy
    calibration.}
    \label{fig:muon_calib_match}
\end{figure}

The muon calibration point was determined by simulating the energy deposited by cosmic-ray muons
in the BC-501A detector using \textsc{geant4}, following Ref.~\cite{zhang_soudan}, and using the energy and angular distributions from Ref.~\cite{dakota2}. The
simulated distribution has the Landau-like shape expected for ionization energy loss in a finite
detector. After accounting for the finite detector resolution, the simulated and measured
pulse-height spectra were compared by matching both the minimum-ionizing peak and its high-energy
tail, as shown in \cref{fig:muon_calib_match}. This procedure gives a most-probable voltage signal of \SI{2.423}{\volt} and constrains the high-energy calibration, where the scintillator
response is approximately linear.

\begin{figure}[htb]
    \centering
    \includegraphics[width=\linewidth]{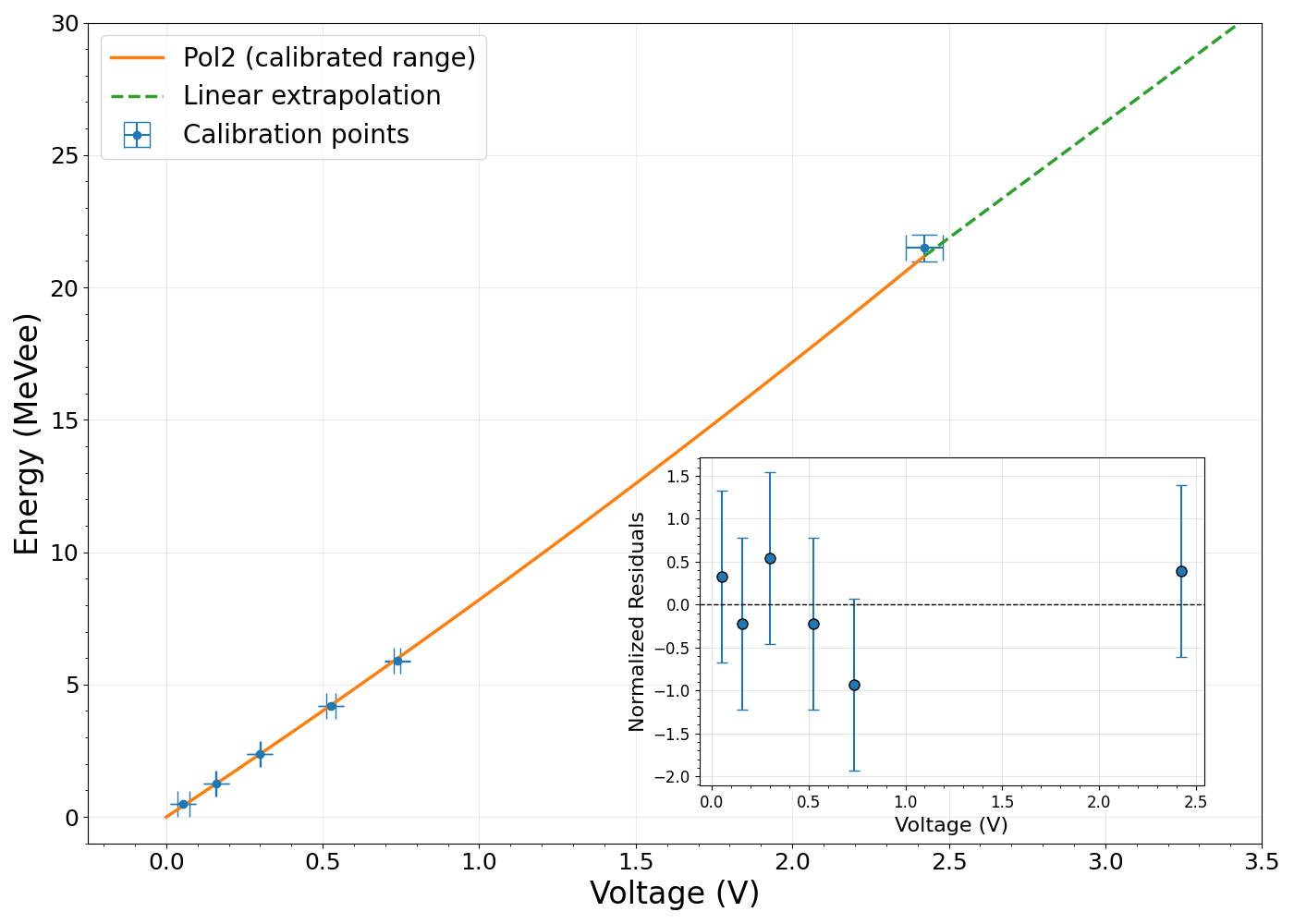}
    \caption{BC-501A energy calibration. The low-energy response is described by a
    second-order polynomial constrained by the source Compton edges and the muon calibration
    point, while the high-energy response is described by a linear continuation. The inset shows
    the normalized residuals between the calibration data and the polynomial fit.}
    \label{fig:calibration_curve}
\end{figure}

The resulting calibration is described by a second-order polynomial below the muon calibration
point and a linear relation at higher energies:
\begin{equation}
    E(V)=
    \begin{cases}
        aV^2+bV+c, & V<\SI{2.423}{\volt}, \\[2pt]
        8.746\,V, & V\geq\SI{2.423}{\volt},
    \end{cases}
    \label{eq:bc501-energy-calibration}
\end{equation}
where $a=(0.379\pm0.162)$~\SI{}{MeVee\per\square\volt}, $b=(7.832\pm0.151)$~\SI{}{MeVee\per\volt}, and
$c=-0.008\pm0.025$~\SI{}{MeVee}. The quoted parameter uncertainties are obtained
from the diagonal elements of the fit covariance matrix using the
absolute input uncertainties. At the transition voltage
$V_{\mu}=\SI{2.423}{\volt}$, the fitted polynomial gives
$E(V_{\mu})\approx21.191$~MeVee. The high-energy linear slope is
defined as $E(V_{\mu})/V_{\mu}$, ensuring continuity between the
two branches. The simulated muon energy listed in
\cref{tab:calibration-sources} is a fit input, through which the
polynomial is not required to pass exactly. The calibration curve
is shown in \cref{fig:calibration_curve}.

\subsection{Underground Measurements}\label{sec:bc501-underground}

The muon-induced neutron measurement was performed in the Subatomic
Particle Hideout with the BC-501A detector resting on its side,
with its cylinder axis horizontal. The detector was positioned
approximately \SI{20}{\centi\meter} above the floor,
\SI{50}{\centi\meter} from the south-facing wall, and
\SI{140}{\centi\meter} from the laboratory door.
Data were acquired for a total live time of
$T_{\rm live}=\SI{1431}{\hour}$ using a hardware trigger threshold
of \SI{300}{\milli\volt}. The waveform of each triggered event
was recorded for offline analysis.

\begin{figure}[htb]
    \centering
    \includegraphics[width=\linewidth]{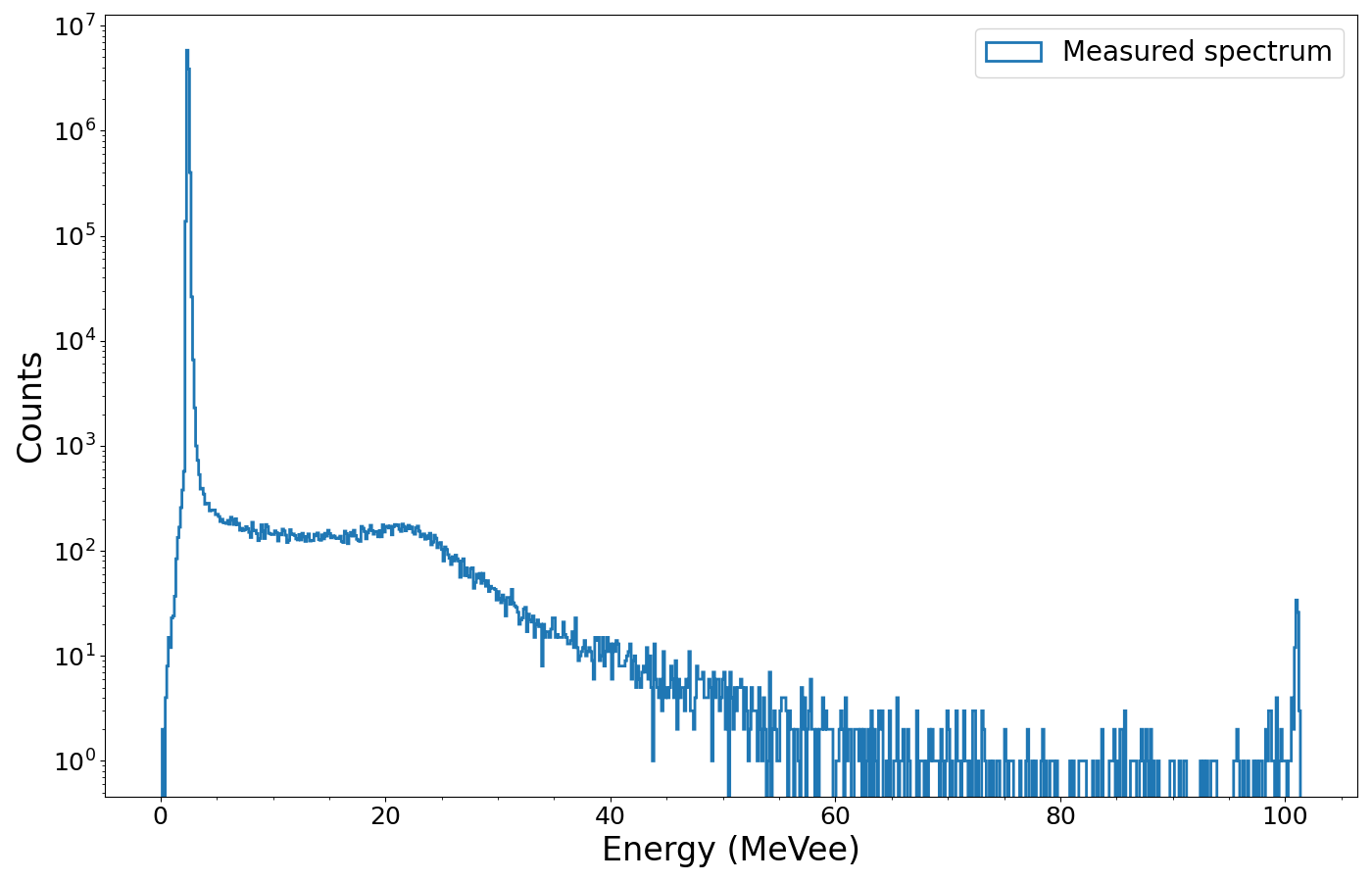}
    \caption{Pulse-height spectrum recorded with the BC-501A detector in the Subatomic Particle
    Hideout. The measured pulse amplitudes are converted to visible energy using
    Eq.~\ref{eq:bc501-energy-calibration}.}
    \label{fig:sph-bc501-pulse-height}
\end{figure}

\begin{figure}[htb]
    \centering
    \includegraphics[width=\linewidth]{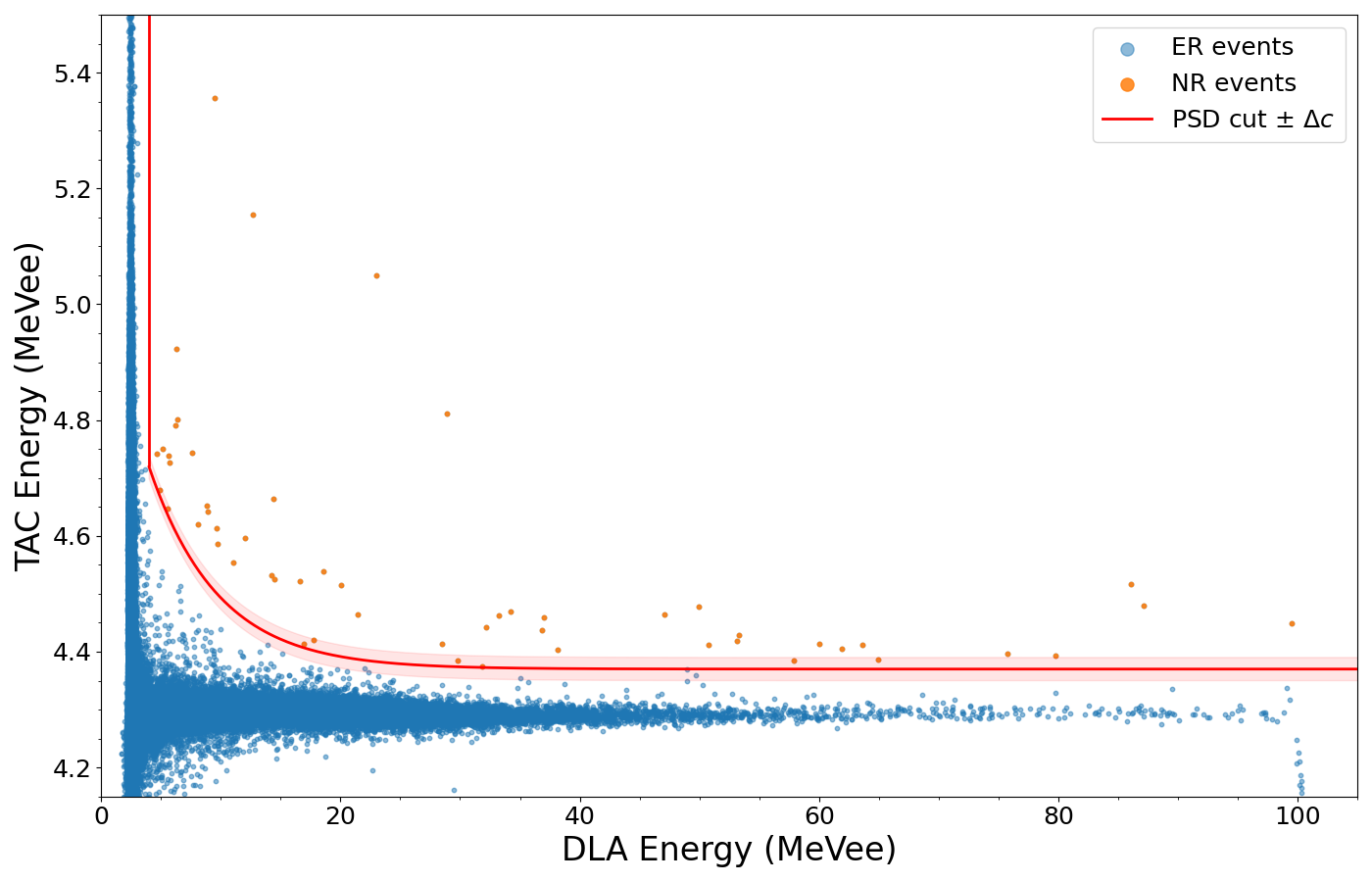}
\caption{Pulse shape discrimination applied to the underground
BC-501A data. The solid red line shows the nominal selection
boundary, and the shaded region shows the offset variation
$c$ within the $\Delta c$ range, used to evaluate the sensitivity of the selected
count. Events above the curve and satisfying the vertical
pulse-amplitude requirement are classified as neutron candidates.}
    \label{fig:sph-psd-cut}
\end{figure}

The pulse-height spectrum of all recorded events is shown in
\cref{fig:sph-bc501-pulse-height}, with the measured voltage converted to visible energy using
Eq.~\ref{eq:bc501-energy-calibration}. The minimum-ionizing muon peak is visible near \SI{21}{MeVee}. Applying the pulse shape analysis described in
Sec.~\ref{sec:psd-procedure} separates the data into the populations shown in
\cref{fig:sph-psd-cut}. The dense electronic-recoil band contains events induced primarily by
$\gamma$ rays and muons, while the less populated nuclear-recoil band contains the neutron
candidates. The two populations are well separated at high visible energy but converge toward
the low-energy region, where the discrimination deteriorates.

Neutron candidates were selected using an empirical,
energy-dependent boundary in TAC amplitude. The boundary was
chosen manually to follow the separation between the observed
electronic- and nuclear-recoil populations while suppressing
electronic-recoil contamination. Its functional form is
\begin{equation}
    f(E)=a\,e^{-E/E_0}+c\,,
    \label{eq:psd-cut}
\end{equation}
where $E$ is the visible energy, $a=0.70$,
$E_0=5.75$~MeVee, and $c=4.37$, with $a$ and $c$ expressed
in the TAC-amplitude units used in \cref{fig:sph-psd-cut}.
The exponential term raises the selection boundary at low
visible energy, where the populations approach one another,
while the constant term sets the high-energy boundary.
Events above this boundary and satisfying the pulse-amplitude
requirement are classified as neutron candidates.

%A vertical selection boundary is additionally imposed to exclude the low-energy region where the
%two populations cannot be reliably separated. This boundary corresponds to a voltage of \SI{0.577}{\volt} or a visible energy of $4.63\pm0.17$~MeVee, where the uncertainty is propagated from
%the energy calibration. This value defines the analysis threshold, such that it is the minimum recorded
%visible signal that can satisfy both the energy and pulse shape requirements for classification as
%a neutron-induced event.
A vertical selection boundary at a measured pulse amplitude of
\SI{0.577}{\volt} excludes the low-energy region where the nuclear-
and electronic-recoil populations cannot be reliably separated.
Using Eq.~\ref{eq:bc501-energy-calibration}, this boundary corresponds
to a visible-energy threshold of $(4.63 \pm 0.17)$~MeVee. The
calibration uncertainty is assigned conservatively by varying each
coefficient within its quoted uncertainty interval and taking the
largest absolute deviation from the nominal calibrated energy over
all parameter combinations. Neutron candidates must satisfy both
this pulse-amplitude requirement and the PSD boundary. The selection
is applied to individual events using their recorded pulse amplitudes,
without histogram binning.

The sensitivity of the selected count to the PSD boundary
was evaluated by varying $c$ within an offset-variation range $\Delta c = 0.02$,
while holding the shape parameters and vertical pulse-amplitude
boundary fixed. This variation includes or excludes events
near the nominal boundary and gives
\begin{equation}
    N_n = 55 \pm 7_{\rm stat}\,{}^{+5}_{-8\,\rm sys}
\end{equation}
neutron candidates over the \SI{1431}{\hour} live time.
The statistical uncertainty follows from Poisson counting
statistics, while the quoted systematic range describes
the dependence on the adopted PSD boundary.

\subsection{PSD Event Corrections}

The pulse-shape discrimination selection introduces two
potential sources of bias in the measured neutron count:
electronic-recoil events may be misidentified as nuclear
recoils, while genuine neutron-induced recoil events
may be excluded due to the finite neutron acceptance of the
PSD selection. Corrections for both effects were therefore
studied for the measured neutron flux.

To estimate ER contamination, the DLA pulse-height
distribution recorded during the underground measurement
was used. The low-energy portion of this spectrum is
dominated by the radiogenic ER background, providing
an ER-enhanced sample from which the probability of
misidentification can be evaluated. Events satisfying
a horizontal TAC threshold of \SI{4.37}{MeVee} were
selected. This threshold corresponds to the constant
offset of the nominal PSD boundary and conservatively
includes events that would otherwise be rejected by
the energy-dependent component.

\begin{figure}[!t]
    \centering
    \includegraphics[width=\linewidth]
    {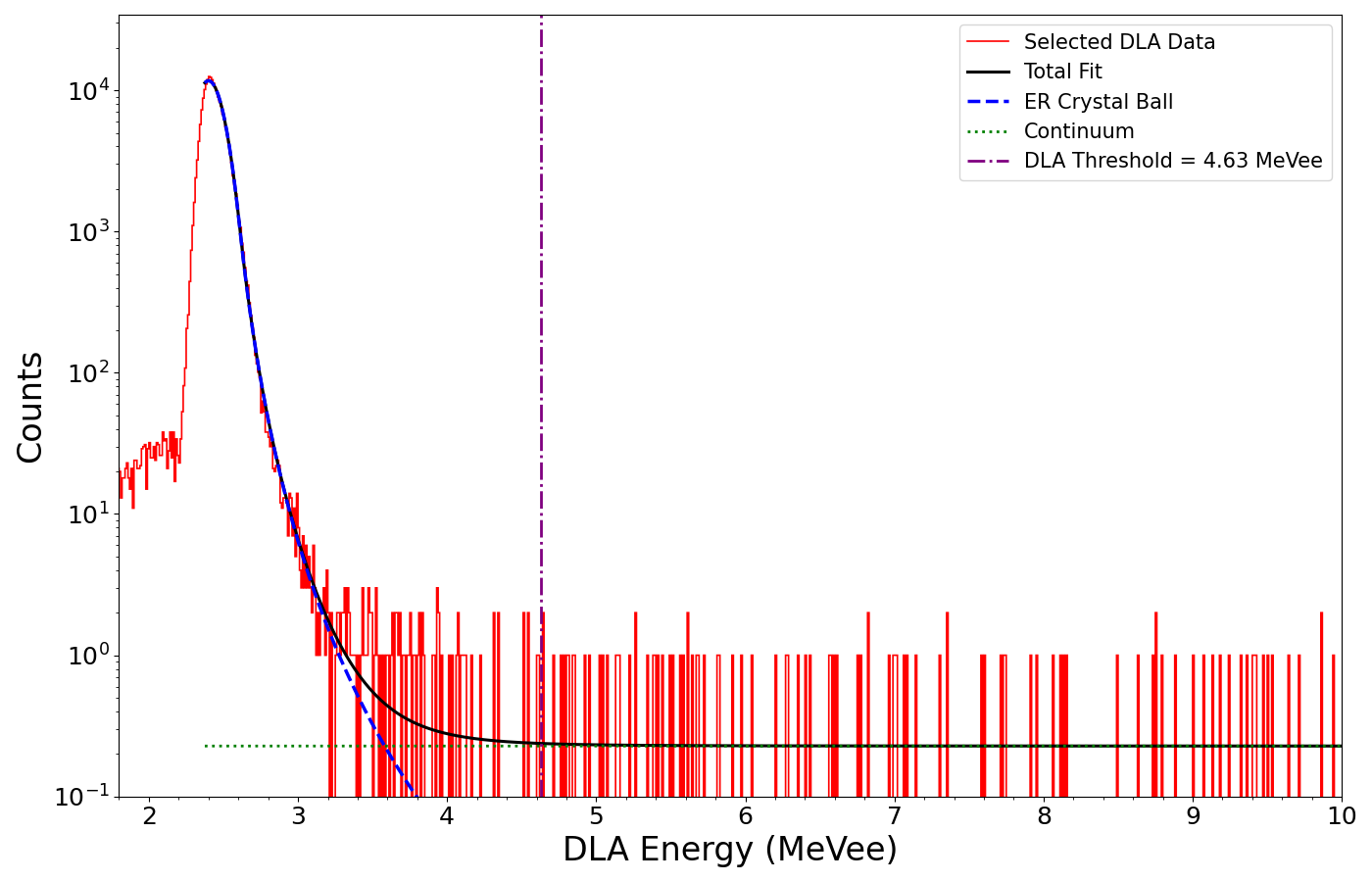}
    \caption{Crystal Ball fit to the low-energy DLA
    pulse-height distribution of the ER-dominated
    sample. The vertical line indicates the
    \protect\SI{4.63}{MeVee} DLA threshold used in the
    PSD selection.}
    \label{fig:crystal-ball-ER-fit}
\end{figure}

The resulting distribution was fitted with a Crystal
Ball function~\cite{gaiser1982crystalball,
oreglia1980crystalball} and an additional continuum
component. The Crystal Ball function describes the
dominant ER peak, while the continuum accounts for
events outside this peak, which may include genuine
neutron-induced recoils. Both components were included
in the total fit, but only the Crystal Ball contribution
was used to estimate ER leakage, since the continuum
cannot be uniquely attributed to ER events. The fit,
shown in Fig.~\ref{fig:crystal-ball-ER-fit}, was
performed by minimizing the Poisson negative
log-likelihood, yielding a reduced Poisson deviance
of $D/\mathrm{ndof}=1.1$, where $D$ is the Poisson
deviance.

The expected number of events from the fitted ER
peak above the DLA threshold was estimated by
integrating the Crystal Ball component,
\begin{equation}
    N_{\rm ER}
    = N_{\rm CB}
    \int_{E_{\rm cut}}^{E_{\rm max}}
    p_{\rm CB}(E)\,dE,
    \label{eq:er-leakage}
\end{equation}
where $N_{\rm CB}$ is the fitted peak normalization,
$p_{\rm CB}(E)$ is the normalized Crystal Ball
distribution, and $E_{\rm cut}=4.63$~MeVee is the
DLA analysis threshold.

This yields an estimated ER contribution of
$N_{\rm ER}=1.01\pm0.32_{\rm stat.}\,^{+1.08}_{-0.47}\,_{\rm sys.}$ events, where
the quoted systematic uncertainty was obtained by
varying the lower bound of the fit interval, and the statistical uncertainty is calculated from the correlations in the fitting parameters. The corresponding multiplicative correction to the
measured neutron count is
\begin{equation}
    \begin{split}
    C_{\rm ER}
    &=1-\frac{N_{\rm ER}}{N_{\rm meas}}\\
    &=0.98\pm 0.01_{\mathrm{stat.}}\,^{+0.01}_{-0.02}\, _{\mathrm{sys.}},
    \end{split}
    \label{eq:er-corr}
\end{equation}
with $N_{\rm meas}$ being the 55 neutron candidates.

To estimate the neutron acceptance of the PSD
selection, a second data set was acquired at the
surface in a laboratory adjacent to the room housing
the plutonium-beryllium source described in
Sec.~\ref{sec:pube-source}. The measurement used
the same detector and acquisition settings as the
underground measurement, including the amplifier
gain and acquisition threshold. The proximity to
the PuBe source provided a substantially larger
population of neutron-induced recoil events than
the underground measurement, particularly at low
visible energies where the ER and neutron
populations approach one another and PSD losses
are most likely to occur.

\begin{figure}[!h]
    \centering
    \includegraphics[width=\linewidth]
    {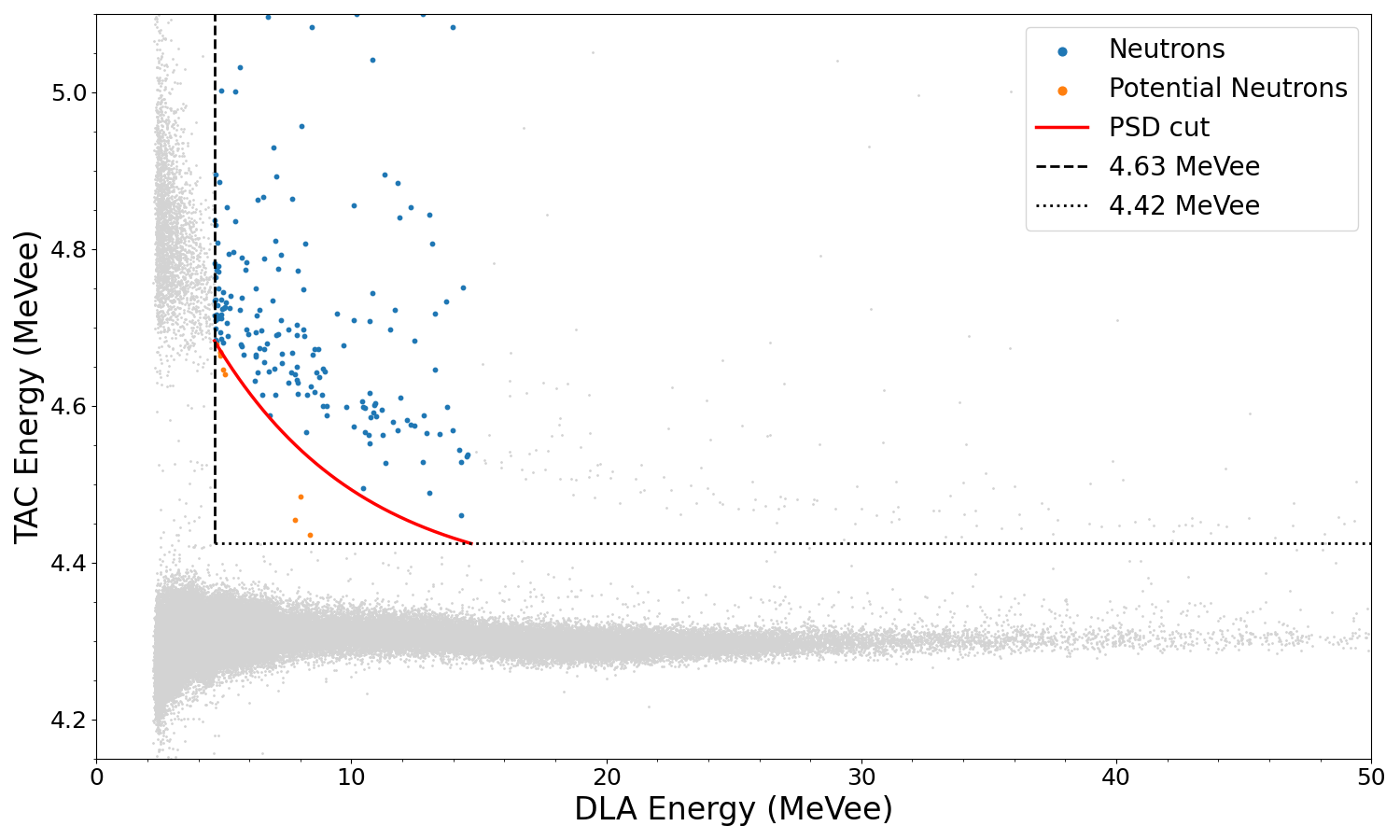}
    \caption{PSD selection applied to the PuBe calibration
    data. The gray points represent the recorded event
    population, while the blue and orange points indicate
    neutron candidates accepted and rejected, respectively,
    by the nominal PSD boundary (red curve) within the
    neutron-enriched control region. The vertical dashed
    line indicates the \SI{4.63}{MeVee} DLA threshold,
    and the horizontal dotted line indicates the
    additional TAC threshold used to suppress ER contamination.}
    \label{fig:neutron-losses}
\end{figure}

The nominal PSD boundary was applied to this
data set, as shown in Fig.~\ref{fig:neutron-losses}.
To suppress ER contamination, an additional
horizontal TAC threshold of \SI{4.42}{MeVee} was
imposed, together with the \SI{4.63}{MeVee} DLA
threshold. The TAC threshold was selected to
sufficiently suppress the ER population while
remaining representative of the low-energy region
of the underground data, where the exponential
PSD boundary approaches its constant offset of
\SI{4.37}{MeVee}. The upper DLA energy limit,
approximately \SI{14.7}{MeVee}, was defined
by the intersection of the horizontal threshold
with the nominal PSD boundary. This restricts
the analysis to the energy interval in which
events passing the horizontal threshold may
still be rejected by the nominal PSD selection,
thereby defining a neutron-enriched control
region.

Within this region, 199 candidate neutron events
were identified, of which six fell below the
nominal PSD boundary. The corresponding neutron
acceptance is
\begin{equation}
    \begin{split}
    \epsilon_{\rm PSD} = \frac{N_{\rm accepted}}{N_{\rm total}}\times100
    &= (96.98\pm1.21_{\rm stat})\,\%,
    \end{split}
    \label{eq:psd-nr-acceptance}
\end{equation}
where the statistical uncertainty is obtained from
binomial counting statistics.

The sensitivity to the control-region definition
was evaluated using two alternative selections
chosen to probe limiting regions of the PSD
boundary. The first restricted the analysis to
the low-energy interval
$4.63 < E_{\rm DLA} < 10.6$~MeVee, with a
more restrictive TAC threshold of
\SI{4.48}{MeVee}, emphasizing the region where
the exponential component of the PSD boundary
is most significant. The second considered
$E_{\rm DLA}>14.7$~MeVee without an upper
energy limit, using a TAC threshold of
\SI{4.37}{MeVee}, corresponding to the constant
offset approached by the PSD boundary at high
energies. These selections yielded acceptances
of $(96.73\pm1.44_{\rm stat})\,\%$ and
$(96.70\pm0.98_{\rm stat})\,\%$, respectively.
The maximum deviation from the nominal acceptance
was 0.28 percentage points, substantially smaller
than the statistical uncertainty, and was therefore
neglected as an additional systematic contribution.

The resulting correction for neutron events excluded
by the PSD selection is

\begin{equation}
    \begin{split}
    C_{\rm NR}
    =\frac{1}{\epsilon_{\rm PSD}}
    =1.031\pm0.013_{\rm stat}\,.
    \end{split}
    \label{eq:nr-corr}
\end{equation}
This correction accounts for the finite neutron
acceptance of the PSD selection, which is not included
in the simulated detector efficiency (Sec.~\ref{sec:bc501-efficiency}) defined by the
visible-energy threshold alone.

\subsection{Detector Efficiency}\label{sec:bc501-efficiency}

The BC-501A detection efficiency depends on both neutron energy and arrival direction. An incident
neutron must interact in the scintillator and deposit sufficient visible energy to exceed the
\SI{4.63}{MeVee} analysis threshold. The efficiency must therefore be averaged over the energy
and angular distributions of the incident neutron field. We determine this flux-weighted efficiency
using a dedicated \textsc{geant4} simulation of the detector in the SPH, with neutrons sampled
from the muon-induced spectrum and angular distribution predicted in Ref.~\cite{dakota2} at the rock--cavern boundary.

The visible energy produced by a nuclear recoil is reduced relative to that produced by an
electron with the same deposited energy. The deposited energies of the individual recoil species
were therefore converted to MeVee using species-dependent light-output functions for protons,
$\alpha$ particles, and carbon, beryllium, and boron recoils~\cite{CECIL1979439,
bc501-light-output}:
\begin{align}
L(E_p)
    &=0.83E_p-2.82\left(1-e^{-0.25E_p^{0.93}}\right),
    \label{eq:lo-proton}\\
L(E_\alpha)
    &=0.41E_\alpha-0.59\left(1-e^{-0.065E_\alpha^{1.01}}\right),
    \label{eq:lo-alpha}\\
L(E_{\mathrm C})
    &=0.017E_{\mathrm C},
    \label{eq:lo-carbon}\\
L(E_{\mathrm{Be}})
    &=0.0821E_{\mathrm{Be}},
    \label{eq:lo-beryllium}\\
L(E_{\mathrm B})
    &=0.0375E_{\mathrm B}.
    \label{eq:lo-boron}
\end{align}
Here, the deposited recoil energies are expressed in \si{\MeV} and the corresponding light
outputs in MeVee. The visible contributions from all recoil products in an event were summed,
while energy deposited through electromagnetic interactions was treated on the electron-equivalent
scale.

Near the analysis threshold, the deposited energy is dominated by proton and carbon elastic
recoils. Carbon-inelastic reactions, including
$^{12}\mathrm{C}(n,\alpha)^{9}\mathrm{Be}$ and
$^{12}\mathrm{C}(n,p)^{12}\mathrm{B}$, nevertheless contribute at the few-percent level.
Including the light-output response of all relevant recoil species therefore provides a more
complete reconstruction of the detector signal.

%The reconstructed visible energy was then compared with the incident neutron energy, as shown in
%\cref{fig:neutron-threshold}. Events satisfying the \SI{4.63}{MeVee} analysis threshold originate
%predominantly from incident neutrons above
%\begin{equation}
%    E_{\rm thr}=10.5\pm0.4~\si{\MeV}.
%\end{equation}
%We therefore define \SI{10.5}{\MeV} as the operational incident-neutron threshold of the
%measurement. This is not a strict kinematic boundary since a very small number of lower-energy neutrons can
%produce signals above the visible-energy threshold, while some higher-energy neutrons deposit
%insufficient energy to be detected. The quoted uncertainty accounts for the finite width of the
%simulated transition and the propagated uncertainty in the \SI{4.63}{MeVee} analysis threshold.
%The measurement consequently constrains the integral flux
%$\Phi_n(E_n>\SI{10.5}{\MeV})$ rather than reconstructing the differential neutron spectrum.

The operational incident-neutron threshold was defined using simulated
events in the lowest visible-energy band,
$4.63 \leq E_{\rm vis}<8.0$~MeVee. The fifth percentile of the
corresponding incident-neutron energy distribution was adopted as
the reporting threshold, giving
\begin{equation}
    E_{\rm thr}=(10.5\pm0.4)~\si{\MeV}.
\end{equation}
Approximately \SI{95}{\percent} of the simulated events in this band
therefore originate from neutrons above the reporting threshold.
Its systematic uncertainty was evaluated by varying the visible-energy
threshold over $(4.63\pm0.17)$~MeVee and varying the percentile
criterion from the third to the seventh percentile. The contributions
from these two variations were combined in quadrature.

Across all simulated events satisfying
$E_{\rm vis}\geq4.63$~MeVee, a fraction
$f_{<}=0.01574$ originates from neutrons with
$E_n<\SI{10.5}{\MeV}$. The reporting threshold is therefore an
operational definition based on the assumed incident spectrum and
detector response, rather than a strict kinematic boundary.
This below-threshold contribution is accounted for when converting
the measured candidate count into the integral flux above
\SI{10.5}{\MeV}, as described in Eq.~\ref{eq:mu-n-flux}.

\begin{figure}[htb]
    \centering
    \includegraphics[width=\linewidth]{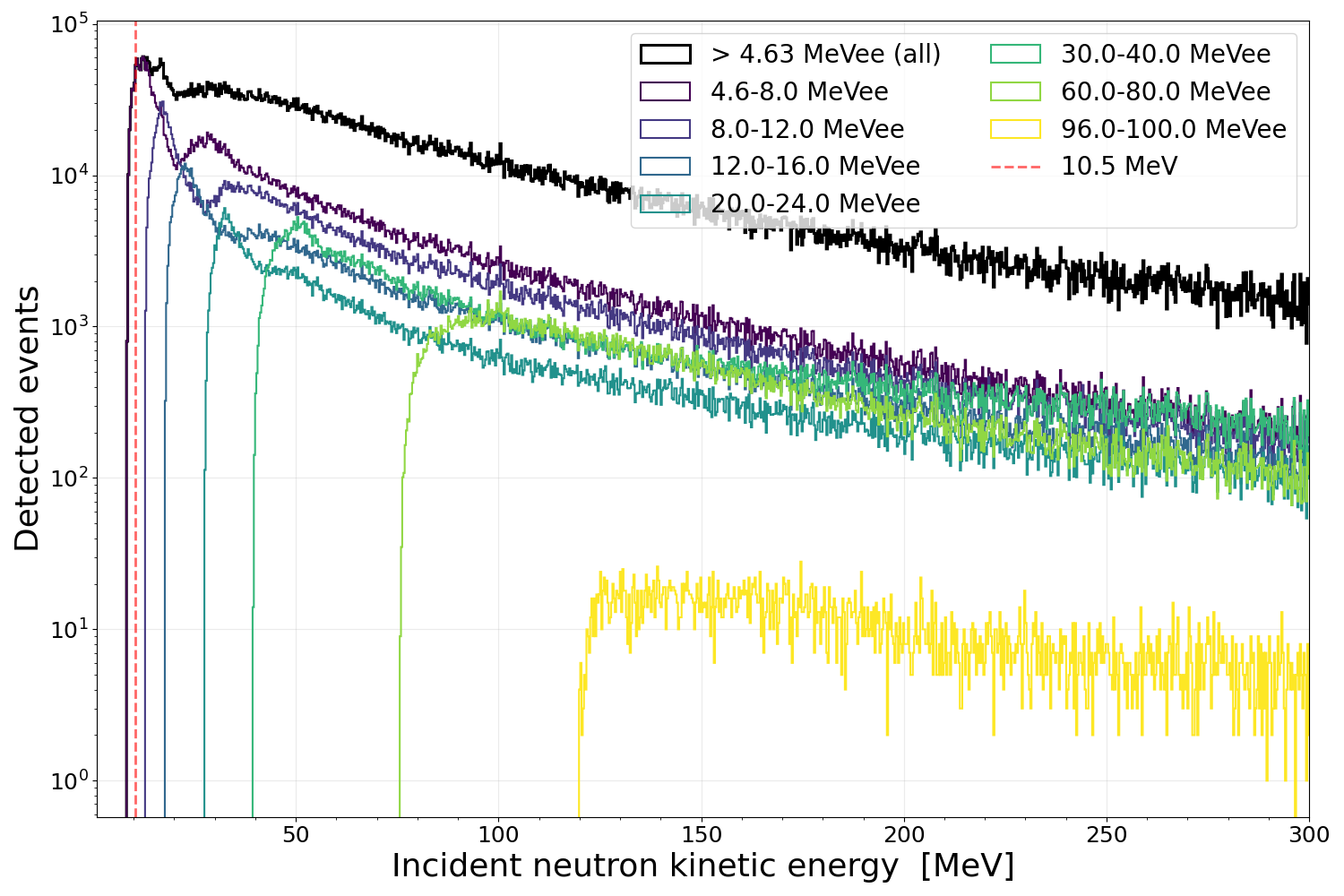}
    \caption{Simulated relationship between neutron energy at the rock--cavern boundary and
reconstructed visible energy in the BC-501A detector. Events satisfying the
\SI{4.63}{MeVee} analysis threshold originate predominantly from neutrons above the
operational boundary-energy threshold of \SI{10.5}{\MeV}, indicated by the vertical line.}
    \label{fig:neutron-threshold}
\end{figure}

The flux-weighted efficiency was calculated for neutrons with boundary energy
$E_n\geq E_{\rm thr}$ that enter the detector. In each boundary-energy bin,
$N_{\rm inc}(E_k)$ is the number of these neutrons entering the detector and
$N_{\rm det}(E_k)$ is the number producing a reconstructed visible signal at or above
\SI{4.63}{MeVee}. Because the source neutrons are sampled from the predicted
boundary spectrum and transported to the detector, the flux-weighted efficiency is
\begin{equation}
    \epsilon
    =
    \frac{\displaystyle\sum_k N_{\rm det}(E_k)}
         {\displaystyle\sum_k N_{\rm inc}(E_k)},
    \qquad E_k\geq E_{\rm thr}.
    \label{eq:weighted-efficiency}
\end{equation}
This gives
\begin{equation}
    \epsilon=7.62^{+1.30}_{-1.99}\,\%.
\end{equation}
The quoted uncertainty is systematic. It was evaluated by varying the visible-energy threshold
over $4.63\pm0.17$~MeVee and repeating the simulated mapping to determine the corresponding
variation in $E_{\rm thr}$. The efficiency was then recalculated using each paired
visible-energy and incident-neutron threshold. This procedure preserves the correlation between
the two quantities and avoids treating the same threshold uncertainty as two independent
contributions.

\subsection{Effective Area and Predicted Rate}\label{sec:mu-n-sph-sims}

Converting the measured neutron count into a flux at the rock--cavern
boundary requires the detector acceptance and intrinsic efficiency.
These quantities were evaluated using a downstream \textsc{geant4}
version~11.2.1 simulation with the \textsc{shielding\_hp} physics list
and G4NDL neutron data. Each source neutron was initialized with the
position, energy, and direction of its corresponding recorded crossing
of the rock--cavern boundary in Ref.~\cite{dakota2}, preserving the
correlations among these quantities. The surrounding rock geometry
and air-filled SPH cavern were retained, allowing subsequent neutron
propagation and scattering within the modeled geometry. A detailed
BC-501A detector model, constructed from the measured dimensions and
manufacturer-specified scintillator
composition~\cite{luxium_bc501a_datasheet}, was placed at the
experimental measurement position. The resulting detector response
therefore includes transport from the cavern boundary to the
scintillator.

The effective area characterizes the fraction of neutrons entering the laboratory that intersect
the detector. It is distinct from the intrinsic efficiency calculated in
Sec.~\ref{sec:bc501-efficiency}, which gives the probability that a neutron entering the detector
produces a signal above the analysis threshold. Neutrons with $E_n\geq E_{\rm thr}$ were generated
at the laboratory boundary over an area
$A_{\rm gen}=\SI{66.185}{\square\metre}$, corresponding to the area used to define the neutron
flux in Ref.~\cite{dakota2}. The effective area relates the boundary flux to the rate of neutrons entering the
detector. For source neutrons with boundary energy $E_n\geq E_{\rm thr}$,
\begin{equation}
    A_{\rm eff}
    =\frac{N_{\rm hit}}{N_{\rm gen}}A_{\rm gen}\,,
    \label{eq:aeff}
\end{equation}
where $N_{\rm gen}$ is the number of neutrons generated at the laboratory boundary and
$N_{\rm hit}$ is the number entering the BC-501A detector volume. This gives
\begin{equation}
    A_{\rm eff}
    =665.29^{+96.05}_{-25.98}~\si{\square\centi\metre}.
\end{equation}

The uncertainty in $A_{\rm eff}$ includes contributions from the incident-neutron threshold, the
detector dimensions, and the reconstructed detector position. The threshold contribution was
evaluated using the correlated visible-energy and incident-neutron threshold variations described
in Sec.~\ref{sec:bc501-efficiency}. The detector height and diameter were each varied by
$\pm\SI{3}{\milli\metre}$, and its position within the SPH was varied by
$\pm\SI{5}{\centi\metre}$ in each coordinate. The position dependence was evaluated using all
eight combinations of these coordinate displacements, with the largest deviations from the nominal
result defining the asymmetric bounds. The position uncertainty dominates the total uncertainty
in $A_{\rm eff}$.

The same detector simulation was used to predict the number of
selected events originating from neutrons with boundary energy
$E_n\geq\SI{10.5}{\MeV}$. Neutrons were generated according to
the flux predicted in Ref.~\cite{dakota2} for an exposure equivalent
to the \SI{1431}{\hour} experimental live time. Applying the
\SI{4.63}{MeVee} analysis threshold gives
\begin{equation}
    N_{\rm sim}=48\pm7_{\rm stat}\pm7_{\rm sys}
\end{equation}
predicted events. The statistical uncertainty is the Poisson uncertainty in the simulated event
count, while the systematic uncertainty arises from the normalization uncertainty of the predicted
neutron flux.

The measured integral neutron flux at the rock--cavern boundary is obtained from
\begin{equation}
    \Phi_{\rm \tiny{RC}}^{\rm exp}(E_n>\SI{10.5}{\MeV})
    =
    \frac{(1-f_{<})C_{\rm ER}C_{\rm NR}N_{\rm meas}}
         {A_{\rm eff}T_{\rm live}\epsilon}\,,
    \label{eq:mu-n-flux}
\end{equation}
where $N_{\rm meas}$ is the observed neutron-candidate count,
$T_{\rm live}$ is the measurement live time, $f_{<}$ is the
simulated fraction of selected events originating below the
incident-neutron reporting threshold, and $C_{\rm ER}$ and $C_{\rm NR}$ are the electronic and nuclear recoil corrections, respectively. The product
$A_{\rm eff}\epsilon$ is the response area calculated for neutrons
above that threshold. The factor $(1-f_{<})$ accounts for the
below-threshold contribution to the selected sample without
changing the reported number of observed candidates.

The visible-energy and incident-neutron thresholds are varied
coherently within each detector-response calculation. For the final flux uncertainty, all four
limiting combinations of the resulting $A_{\rm eff}$ and $\epsilon$ values are evaluated, and the
largest upward and downward deviations are retained as a deliberately conservative uncertainty. The independent geometry-only contribution is then added in quadrature.

\subsection{Results}\label{sec:sph-mu-n-exp}

Applying Eq.~\ref{eq:mu-n-flux} to the
$N_{\rm meas}=55\pm7_{\rm stat}\,{}^{+5}_{-8\,\rm sys}$ neutron candidates recorded over
\SI{1431}{\hour} gives the integral rock--cavern boundary flux of
\begin{equation}
\begin{split}
\Phi_{\rm \tiny{RC}}^{\rm exp}(>\SI{10.5}{\MeV})
&=
\left(
2.09\pm0.29_{\rm stat}\,{}^{+0.79}_{-0.53\,\rm sys}
\right)\\
&\quad\times10^{-7}\,
\si{\per\square\centi\metre\per\second}.
\end{split}
\end{equation}
The statistical uncertainty follows from the Poisson uncertainty
on $N_{\rm meas}$, as well as from the NR correction factor. The systematic uncertainty combines the
variation in the number of candidates selected by the PSD
boundary, the geometry-related uncertainty in $A_{\rm eff}$, the conservative threshold envelope described above, and the ER and NR corrections. These contributions are added in quadrature. The resulting systematic bounds describe a conservative uncertainty range.

The radiogenic contribution to the selected sample was estimated
using the unfolded neutron flux above approximately \SI{7}{\MeV}
and the BC-501A response over the \SI{1431}{\hour} exposure.
This gives approximately 0.1 events before any additional
rejection by the PSD boundary, corresponding to about
\SI{0.2}{\percent} of the 55 selected candidates. This estimate
is small compared with the statistical uncertainty of the
candidate count, and no radiogenic-background subtraction
is applied.

The primary comparison of the measurement is with the integral muon-induced neutron flux predicted in
Ref.~\cite{dakota2},
%\begin{equation}
%    \Phi_n^{\rm pred}(>\SI{10.5}{\MeV})
%    =
%    (1.86\pm0.29_{\rm sys})
%    \times10^{-7}\,
%    \si{\per\square\centi\metre\per\second}.
%\end{equation}
\begin{equation}
\begin{split}
\Phi_{\rm \tiny{RC}}^{\rm pred}(>\SI{10.5}{\MeV})
&=
\left(
1.86\pm0.29_{\rm sys}
\right)\\
&\quad\times10^{-7}\,
\si{\per\square\centi\metre\per\second}.
\end{split}
\end{equation}
The measured central value is approximately \SI{10}{\percent} higher than the prediction but is
consistent within the reported uncertainties. The simulated neutron energy and angular
distributions enter the effective area, flux-weighted efficiency,
and correction for the below-threshold contribution. This comparison
therefore tests the predicted flux normalization under the assumed
spectral shape and detector-response model.

Propagating the predicted boundary field through the detector-level simulation and
converting the count back to the same boundary-flux convention gives
%\begin{equation}
%    \Phi_n^{\rm sim}(>\SI{10.5}{\MeV})
%    =
%    \left(
%    1.84\pm0.27_{\rm stat}\,{}^{+0.69}_{-0.39\,\rm sys}
%    \right)
%    \times10^{-7}\,
%    \si{\per\square\centi\metre\per\second}.
%\end{equation}
\begin{equation}
\begin{split}
\Phi_{\rm \tiny{RC}}^{\rm sim}(>\SI{10.5}{\MeV})
&=
\left(
1.84\pm0.27_{\rm stat}\,{}^{+0.73}_{-0.46\,\rm sys}
\right)\\
&\quad\times10^{-7}\,
\si{\per\square\centi\metre\per\second}.
\end{split}
\end{equation}
This value agrees with the integrated prediction to within approximately \SI{1}{\percent}.
However, the detector-level result is not statistically independent of either the integrated
prediction or the measurement: its normalization originates from Ref.~\cite{dakota2}, while its
conversion to flux uses the same effective-area and efficiency calculations applied to the data.
It is therefore interpreted as a closure test demonstrating the internal consistency of the
detector-response and flux-extraction procedure. The three flux determinations are summarized in
\cref{tab:mu-n-flux-comparison}.

\begin{table}[htb]
\caption{Integral muon-induced neutron flux above \SI{10.5}{\MeV} at the
SPH rock--cavern boundary. All three determinations use the first-entry,
boundary-area normalization of Ref.~\cite{dakota2}, with neutron energy
defined at cavern entry. The direct simulation is a detector-level closure
test using the same response model applied to the measurement.}
\label{tab:mu-n-flux-comparison}
\renewcommand{\arraystretch}{1.35}
\bgroup
\setlength{\tabcolsep}{8pt}
\begin{tabular}{lc}
Determination
& $\Phi(E_n>\SI{10.5}{\MeV})$
  (\si{\per\square\centi\metre\per\second}) \\
\colrule
Integrated prediction~\cite{dakota2}
& $(1.86\pm0.29_{\rm sys})\times10^{-7}$ \\

Direct simulation
& $(1.84\pm0.27_{\rm stat}\,{}^{+0.73}_{-0.46\,\rm sys})
   \times10^{-7}$ \\

Measurement
& $(2.09\pm0.29_{\rm stat}\,{}^{+0.79}_{-0.53\,\rm sys})
   \times10^{-7}$ \\
\end{tabular}
\egroup
\end{table}

Several independently constrained inputs support this agreement. The \textsc{mute} calculation
reproduces the measured underground muon flux and its angular dependence at
CURIE~\cite{dakota}; the rock composition used in the \textsc{geant4} model was measured by
scanning electron microscopy~\cite{dakota2}; and the simulated muon-induced neutron yield
reproduces its measured dependence on muon energy after application of the empirical multiplicity
correction developed in Ref.~\cite{dakota2}. These checks constrain the incident muon field,
local material composition, and total neutron yield.

\begin{figure}[!t]
    \centering
    \includegraphics[width=\linewidth]{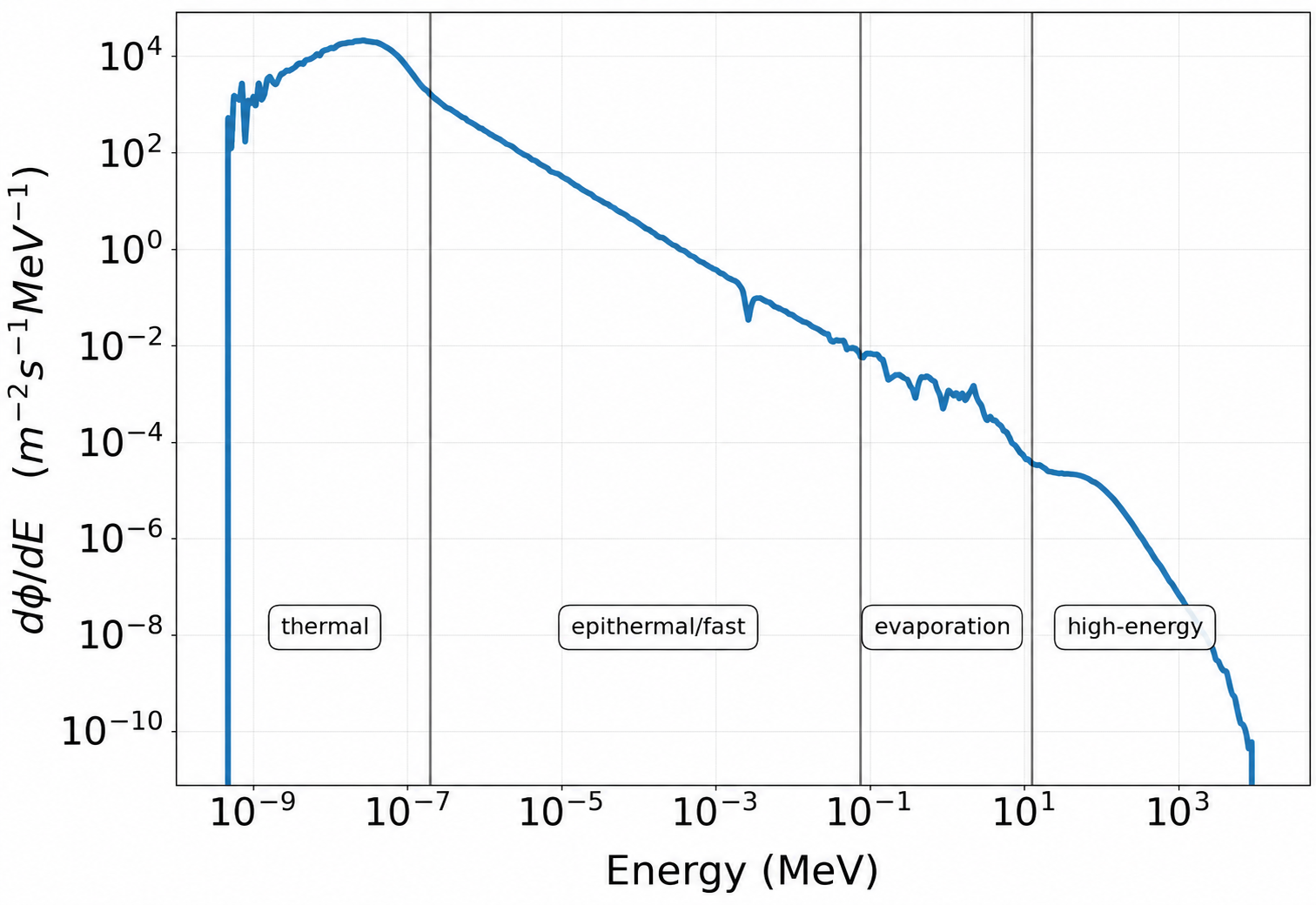}
    \caption{Simulated muon-induced neutron spectrum at the SPH rock--cavern
boundary, scaled by $(1-f_{<})C_{\rm ER}C_{\rm NR}N_{\rm meas}/N_{\rm sim}$. Neutron energy is defined at cavern entry. The measurement constrains
the normalization above \SI{10.5}{\MeV}, while the spectral shape is
taken from Ref.~\cite{dakota2}; the spectrum below this energy is a
model extrapolation.}
    \label{fig:exp-scaled-n-spec}
\end{figure}

The present measurement provides a single integral constraint on the rock--cavern boundary flux above
\SI{10.5}{\MeV} and cannot distinguish a difference in the predicted
normalization from a difference in the neutron spectral shape.
We therefore retain the simulated differential spectrum from
Ref.~\cite{dakota2} and adjust its normalization~\cite{zhang_soudan} using
\begin{equation}
    \frac{d\Phi_{\rm final}}{dE}
    =
    \frac{(1-f_{<})C_{\rm NR}C_{\rm ER}N_{\rm meas}}{N_{\rm sim}}
    \frac{d\Phi_{\rm sim}}{dE}.
    \label{eq:scaled-neutron-spectrum}
\end{equation}
Here, $N_{\rm sim}=48$ includes only selected events originating
from incident neutrons with $E_n\geq\SI{10.5}{\MeV}$, while
$(1-f_{<})C_{\rm NR}C_{\rm ER}N_{\rm meas}$ is the inferred contribution from the
same incident-energy range in the measured sample.
The resulting spectrum, shown in \cref{fig:exp-scaled-n-spec},
retains the simulated boundary spectral shape, with its normalization
constrained by the measured integral flux above \SI{10.5}{\MeV}.
Below this energy, the scaled spectrum represents an extrapolation
of the model.

\begin{table*}[!t]
\caption{Muon-induced neutron fluxes reported at selected underground facilities.
The CURIE result uses the first-entry rock--cavern boundary convention of
Ref.~\cite{dakota2}. Differences in flux definition, detector response, and
neutron-energy threshold limit direct quantitative comparisons.}
\label{tab:urf-mu-n}
\bgroup
\setlength{\tabcolsep}{10pt}
\renewcommand{\arraystretch}{1.35}
\begin{tabular}{lccc}
Facility
& Depth (m.w.e.)
& $\Phi(E_n>E_{\rm thr})$
  (\si{\per\square\centi\metre\per\second})
& $E_{\rm thr}$ (\si{\MeV}) \\
\colrule
Fermilab~\cite{Garrison:2014nla}
& 265
& $(1.5\pm1.4)\times10^{-6}$
& 15 \\

CURIE (this work)
& 406
& $(2.09\pm0.29_{\rm stat}\,{}^{+0.79}_{-0.53\,\rm sys})
   \times10^{-7}$
& 10.5 \\

BUST~\cite{bust_n_flux}
& 850
& $(2.1\pm0.5)\times10^{-8}$
& 28.6 \\

Soudan~\cite{zhang_soudan}
& 2032
& $(2.23\pm0.52_{\rm stat}\pm0.99_{\rm sys})
   \times10^{-9}$
& 20 \\

Boulby~\cite{ARAUJO2008471}
& 2860
& $\approx4.5\times10^{-9}$
& 0.55 \\
\end{tabular}
\egroup
\end{table*}

Direct measurements of the muon-induced neutron flux remain scarce,
particularly near an overburden of \SI{400}{}~m.w.e.
Table~\ref{tab:urf-mu-n} places the present result alongside measurements
at other underground facilities, with their respective depths and
neutron-energy thresholds. Differences in detector response and
energy threshold limit direct quantitative comparisons. The result presented here is, however, qualitatively consistent with these existing measurements, in that the neutron flux decreases approximately with increasing facility depth.

%%%%%%%%%%%%%%%%%%%%%%%%%%%%%%%%%%%%%%%%%%%%%%%%%%%%%%%%%%%%%%%%%%
%%%%%%%%%%%%%%%%%%%%%%%%%%%%%%%%%%%%%%%%%%%%%%%%%%%%%%%%%%%%%%%%%%
\section{Discussion and Conclusions}\label{sec:Discussion}

This work presents the first direct measurements of the radiogenic and muon-induced neutron
backgrounds at the Colorado Underground Research Institute. The two components were measured in
the Subatomic Particle Hideout using complementary detector systems matched to their respective
energy ranges.

The radiogenic component was measured using a custom Bonner Cylinder Spectrometer comprising a
$^{3}$He proportional counter and six cylindrical HDPE moderators. The response functions were
calculated in \textsc{geant4} and calibrated against a $^{239}$Pu-$^{9}$Be source using the
shadow-cone method. Refolding and unfolding the PuBe data provided a closure test of the calibrated
response matrix. To account for the angular dependence of the cylindrical geometry, measurements
at \SI{0}{\degree}, \SI{45}{\degree}, and \SI{90}{\degree} detector rotation were interpolated
and weighted by the simulated angular response and the angular phase space of an isotropic field.
This correction permits the larger active volume of a cylindrical spectrometer to be used without
leaving its directional response unaccounted for.

Applying the corrected response matrix to the seven underground count
rates gives a total low-energy ambient scalar neutron flux at the detector for
$E_n\lesssim\SI{10}{\MeV}$ of
\begin{equation}
    \Phi_n
    =
    (2.51\pm0.58_{\rm stat}\pm0.39_{\rm sys})
    \times10^{-5}\,
    \si{\per\square\centi\meter\per\second}.
\end{equation}
This field is expected to be dominated by radiogenic production;
no cosmogenic contribution has been subtracted. The epithermal
component accounts for approximately \SI{57}{\percent} of the
integrated flux. The six reconstructions obtained using three
unfolding algorithms and two default spectra give consistent broad
thermal, epithermal, and fast components. This consistency supports
stability against the tested changes in algorithm and input
component weights, while the detailed component shapes remain
less constrained. The variation among underground measurements
in \cref{sph-radiogenic-flux}, despite facility depths spanning
\SIrange{140}{6230}{}~m.w.e., underscores the importance of
site-specific measurements: radiogenic neutron production depends
on the local uranium and thorium content and the composition of
the surrounding rock~\cite{mei-alpha-n,alpha-n-depth}.

The muon-induced component was measured with a BC-501A organic liquid scintillator operated for
\SI{1431}{\hour}. Pulse shape discrimination identified 55 neutron candidates above the
\SI{4.63}{MeVee} analysis threshold, corresponding to an operational neutron energy threshold
of \SI{10.5}{\MeV}. After correcting for the effective area, flux-weighted efficiency, the below-threshold contribution, and the electronic and nuclear recoil corrections the
measured integral rock--cavern flux is
%\begin{equation}
%    \Phi_n^{\rm exp}(E_n>\SI{10.5}{\MeV})
%    =
%    \left(
%    2.07\pm0.28_{\rm stat}\,{}^{+0.78}_{-0.52\,\rm sys}
%    \right)
%    \times10^{-7}\,
%    \si{\per\square\centi\metre\per\second}.
%\end{equation}
\begin{equation}
\begin{split}
\Phi_{\rm \tiny{RC}}^{\rm exp}(>\SI{10.5}{\MeV})
&=
\left(
2.09\pm0.29_{\rm stat}\,{}^{+0.79}_{-0.53\,\rm sys}
\right)\\
&\quad\times10^{-7}\,
\si{\per\square\centi\metre\per\second}.
\end{split}
\end{equation}
This central value is approximately \SI{12}{\percent} higher than the
$(1.86\pm0.29_{\rm sys})\times10^{-7}$~\si{\per\square\centi\metre\per\second} prediction of
Ref.~\cite{dakota2} but is consistent within the reported uncertainties. Because the predicted
neutron energy and angular distributions enter the calculations of $A_{\rm eff}$ and $\epsilon$,
the measurement tests the normalization of the predicted flux under the simulated response model;
it does not independently determine the neutron spectral shape. The detector-level simulation
provides an additional closure test but shares inputs and uncertainties with the flux extraction
and is therefore not treated as an independent comparison.

The agreement with Ref.~\cite{dakota2} is supported by several separately constrained inputs. The
\textsc{mute} framework reproduces the measured underground muon flux and its angular dependence
at CURIE~\cite{dakota}; the rock composition used in the \textsc{geant4} model was determined
experimentally; and the simulated dependence of neutron yield on muon energy was tested against
existing measurements before applying the empirical multiplicity correction of
Ref.~\cite{dakota2}. The present result therefore provides an experimental benchmark for the
corrected simulation framework at shallow depth. Such measurements remain valuable because
general-purpose transport calculations can underpredict muon-induced neutron yields by factors of
two or more in some materials and energy regimes~\cite{ARAUJO2005398,REICHHART201367,
sutanto_mu_n}.

Together, the measurements characterize complementary aspects of the neutron
environment at CURIE. The BCS determines the local low-energy scalar flux,
which is expected to be dominated by radiogenic production, while the BC-501A
measurement constrains the high-energy muon-induced flux at the rock--cavern
boundary. Because these quantities use different flux definitions, their
numerical values do not directly establish the relative local contributions
of the two components. Such a comparison would require the predicted
muon-induced field to be evaluated as a local scalar flux at the BCS position.
The measurements are nevertheless consistent with the expected transition
from a low-energy background governed primarily by local radioactivity to a
high-energy background dominated by muon-induced neutrons~\cite{mei_hime,alpha-n-depth}.

The measurements also have important limitations. The radiogenic unfolding is intrinsically
underdetermined because the seven response functions are broad and strongly overlapping.
Consequently, the measurement constrains the relative strengths of the thermal, epithermal, and
fast components more reliably than fine spectral structure. The largest statistical contribution
to the radiogenic flux uncertainty arises from the response-function scaling of the bare
configuration, whose PuBe calibration rate was low. For the muon-induced measurement, the loss of
neutron--$\gamma$ discrimination at low visible energy restricts the analysis to
$E_n>\SI{10.5}{\MeV}$. Furthermore, the 55 selected events provide an integral constraint but
are insufficient to reconstruct the differential muon-induced spectrum. Resolving that spectrum
would require a substantially longer exposure, a larger detector volume, or both.

The methods developed here are applicable to other low-background facilities. The isotropic
correction provides a framework for using cylindrical moderator geometries in low-flux
environments, where increasing the active counter volume can reduce the required acquisition time.
For fast-neutron measurements, the combination of species-dependent light-output reconstruction,
a simulation-derived incident-energy threshold, and a flux-weighted efficiency provides a
consistent method for converting a PSD-selected event count into an integral neutron flux.
Together with the underground muon measurements of Ref.~\cite{dakota} and the muon-induced
background calculations of Ref.~\cite{dakota2}, these results provide a combined characterization
of the radiogenic and cosmogenic backgrounds at CURIE and the quantitative inputs needed to design
and evaluate future low-background experiments at the facility.

\section*{Acknowledgments}\label{sec:Acknowledgments}
This material is based upon work supported by a National Science Foundation Graduate Research Fellowship under Grant No. DGE-2137099, a U.S. Department of Energy Office of Science Grant No. DE-FG02-93ER40789, and the Colorado School of Mines via faculty startup funds and the ARCS Foundation.

%\FloatBarrier
%\clearpage
\bibliography{CURIE_references.bib}

%\printbibliography

\end{document}